\documentclass[a4paper]{elsart}
\usepackage{literat}
\usepackage{cite}
\usepackage{graphicx}
\usepackage{amsmath}
\usepackage{amssymb}
\usepackage{graphicx}
\usepackage{bm}

\begin{document}
\begin{frontmatter}
\title{Entropy driven mechanism for noise induced patterns formation in reaction-diffusion systems}
\author{D.O.Kharchenko,}
\ead{dikh@ipfcentr.sumy.ua}
\author{S.V.Kokhan}
\address{Institute of Applied Physics, Nat. Acad. Sci. of Ukraine, Sumy,  Ukraine}
\author{A.V. Dvornichenko}
\address{Sumy State University, Sumy,  Ukraine}


\begin{abstract}
We have studied the entropy-driven mechanism leading to stationary patterns
formation in stochastic systems with local dynamics and non-Fickian diffusion.
We have shown that a multiplicative noise fulfilling a fluctuation-dissipation
relation is able to induce and sustain stationary structures with its intensity
growth. It was found that at small and large noise intensities the system is
characterized by unstable homogeneous states. At intermediate values of the
noise intensity three types of patterns are possible: nucleation, spinodal
decomposition and stripes with liner defects (dislocations). Our analytical
investigations are verified by computer simulations.
\end{abstract}

\end{frontmatter}

\section{Introduction}

It is well known that the interaction of noise and nonlinearity attracts a lot
of interest in nonequilibrium systems theory \cite{SSGO2007}. Most of articles,
concerning above noise induced phenomena, relate to problems of external noise
influence: noise induced transitions in zero-dimensional systems \cite{Horst};
noise-induced phase transitions \cite{VB94,Garcia}; stochastic resonance
\cite{WM95}; noise sustained patterns \cite{GOHMS93,BK94,PBB96,ZSG98}, etc.
Analytically, numerically and in natural experiments it was found that an
external noise plays an organizing role only if its amplitude depends on the
field variable
--- multiplicative noise (see, for example, Refs.\cite{Horst,Garcia,EPJ2002,Kaw2003}).
If more than one noise source is introduced into the system dynamics, then the
noise cross-correlations start to play a crucial role in ordering processes: a
kind of phase transitions principally changes \cite{EPJB2003,PRE2005}. From a
fundamental viewpoint such effects has a dynamical origin: in a short-time
limit external fluctuations destabilize a disordered homogeneous state. An
analytic description of extended systems with external fluctuations is provided
with an approximately known stationary distribution function.

The possibility of noise induced patterns formation has been studied in last
two decades (see, for example,
Refs.\cite{Mikhailov81,GOSRP92,GOHMS93,BK94,PBB96,ZSG98,Garcia,
BISL2003,MHW2005}). One of the most interesting problem when the noise sustains
patterns is widely discussed recently
\cite{Haeften2004,ISMCS03,LGONSG04,MHW2005,SHBESG06}. In the most of works,
concerning this problem, external fluctuations influence on the dynamics of
systems is considered. Recently a new mechanism for noise induced patterns
formation was discovered. Its origin is in a relaxational dynamics governed by
the field-dependent mobility/kinetic coefficient \cite{Bray}. It was shown that
corresponding self-organization processes are not related to a short-time
instability of a homogeneous/disordered state. Here the principle role is
played by entropy variations caused by the field-dependent mobility
\cite{GO2001,GO2003,LB2004,LB2006,condmat,EPJ2008}. Thus, an ordered behaviour
appears due to the balance between relaxing forces moving the system to the
homogeneous state, and fluctuations dependent on the field variable pulling the
system away from the disordered state. This mechanism belongs to a kind of
entropy driven phase transitions that is an extension of noise induced
unimodal-bimodal transitions in zero-dimensional systems \cite{Horst}. In such
systems a dissipation is related to fluctuations according the fluctuation
dissipation relation. Hence, one gets an internal multiplicative noise with
intensity reduced to the bath temperature. For such class of stochastic systems
the corresponding distribution function, free energy and associated effective
potential are known exactly. Therefore, coherent structures/patterns appeared
in the course of self-organization processes can be analyzed without any
dynamical reference.

In this paper we will study a behaviour of extended stochastic systems when the
noise can induce or sustain spatial patterns according to entropy driven
mechanism. To this end we consider a model of reaction-diffusion system of the
field-dependent mobility with non-Fickian diffusion and internal fluctuations
of the multiplicative character. Reaction-diffusion systems play an important
role in the study of generic spatiotemporal behaviour of far from equilibrium
systems. These models are usually applied to describe an extended systems when
the physical space is divided in to the small cells and dynamics of each cell
is described by a variable type of concentration. The model to be considered
concerns the local dynamics of the variables inside each cell and the transport
phenomena between cells. The local dynamics is determined by so-called chemical
reactions inside each cell. The transport is caused by non-Fickian diffusion.
We will study stationary spatial inhomogeneities with lengths bigger than the
linear dimensions of the cells in such system.

Formally, reaction-diffusion models are described by the equation for the field
$x=x(\mathbf{r},t)$ in the form $\partial_tx=f(x)-\nabla J$, where $f(x)$
stands for the local dynamics, $J$ is the flux for transport phenomena,
$\nabla=\partial_\mathbf{r}$. Considering the non-Fickian diffusion, we exploit
a gradient of molecular interactions potential $U(\mathbf{r})$
($U(\mathbf{r})=-\int u(\mathbf{r}-\mathbf{r}')x(\mathbf{r}'){\rm
d}\mathbf{r}'$, where the spherically symmetric interaction potential
$u(\mathbf{r})$ between molecules separated by a distance $|\mathbf{r}|$ is
introduced). The corresponding force given by the gradient of $U(\mathbf{r})$
governs the transport phenomena. In the case of the small interaction radius
compared to the diffusion length the concentration field $x$ will not vary
significantly within the interaction radius. It allows to approximate the
integral by $\kappa x+\beta\nabla^2 x$, where $\kappa=\int u(\mathbf{r}){\rm
d}\mathbf{r}$, $\beta=(1/2)\int |\mathbf{r}|^2u(\mathbf{r}){\rm d}\mathbf{r}$
($\beta\simeq\kappa r_c^2/2$, where $r_c$ is a correlation radius); $\int
\mathbf{r}u(\mathbf{r}){\rm d}\mathbf{r}=0$ due to symmetrical properties of
the interaction potential \cite{ChemReact}. Therefore, the obtained flux allows
to describe phase separation processes with mutual (lateral) interactions; the
combined model with local dynamics can be used to consider the spatial patterns
induced by the fluctuations of the bath. With the help of such formalism one
can find an explicit form of the stationary distribution functional and study
spatial patterns using a variational principle. Analytical results we verify by
computer simulations.

The paper is organized in the following manner. In the Section II we introduce
a general model for stochastic reaction-diffusion systems and present the
theoretical approach is useful in our consideration. In Section III we discuss
the model and assumptions related to our study. Section IV deals with a
theoretical investigation of stationary noise induced/sustained patterns.
Finally we summarize our results in Section V.

\section{Main equations}

Let us start with the typical deterministic equation for the reaction-diffusion
models in the form
\begin{equation}
\partial_t x=f[ x(\mathbf{r},t)]-\nabla J[ x(\mathbf{r},t)],
\end{equation}
where $x$ stands collectively for the variables, $f[x]$ for the local dynamics
inside each cell, $J[x]$ is the flux proportional to the conjugate
thermodynamic force, arising from the spatial variation of the chemical
potential $\mu[ x(\mathbf{r},t)]$. One can exploit a definition $J[
x(\mathbf{r},t)]=-D_{ef}[ x(\mathbf{r},t)]\nabla\mu[ x(\mathbf{r},t)]$, where
the chemical potential $\mu=\delta\mathcal{F}/\delta x$ is introduced through
the free energy $\mathcal{F}[x]$; $D_{ef}[x]$ is an effective diffusion
coefficient related to the field-dependent mobility. In such a case, the
deterministic evolution equation for the field $x$ reads
\begin{equation}\label{eq2}
\partial_t x=f[x]+\nabla\left(D_{ef}[x]\nabla\mu[x]\right).
\end{equation}

Formally, Eq.(\ref{eq2}) can be written in a variational form as
\begin{equation}\label{eq3d}
\partial_t x=-\frac{1}{D_{ef}[x]}\frac{\delta \mathcal{U}[x]}{\delta x},
\end{equation}
where the potential functional $\mathcal{U}[x]$ is represented through the
definitions for $f[x]$, $D_{ef}[x]$ and $\mathcal{F}[x]$. It plays a role of a
Lyapunov functional for the deterministic dynamics. An explicit form for the
functional $\mathcal{U}[x]$ can be found only if $\mu=x$. Indeed, here one has
$$\mathcal{U}[x]=\int{\rm d}{\mathbf{r}}\left\{ -\int f[x']D_{ef}[x']{\rm d
}x'+(D_{ef}[x]\nabla x)^2/2\right\}.$$ Generally, if the chemical potential
$\mu$ is a function of both $x$ and $\nabla x$, then one can obtain only its
first variation,
\begin{equation}\label{dU}
\delta \mathcal{U}[x]=-\int{\rm d}{\mathbf{r}} \delta
x\left\{f[x]D_{ef}[x]+D_{ef}[x]\nabla\left(D_{ef}[x]\nabla\mu[x]\right)\right\},
\end{equation}
which after substitution into Eq.(\ref{eq3d}) gives Eq.(\ref{eq2}) immediately.

Considering the system under real conditions of thermal bath influence, one has
to take into account corresponding fluctuations. In stochastic analysis we
introduce a related multiplicative noise in an \emph{ad hoc} form, following
Ref.\cite{MHW2005},
\begin{equation}\label{eq3}
\partial_t x=-\frac{1}{D_{ef}[x]}\frac{\delta \mathcal{U}[x]}{\delta x}+\frac{1}{\sqrt{D_{ef}[x]}}\xi(\mathbf{r},t).
\end{equation}
where the fluctuation dissipation relation holds, i.e.
$\langle\xi(\mathbf{r},t)\xi(\mathbf{r}',t')\rangle=2\sigma^2\delta(\mathbf{r}-\mathbf{r}')\delta(t-t')$;
$\langle\xi(\mathbf{r},t)\rangle=0$; $\sigma^2$ is the intensity of a Gaussian
noise $\xi$.

Using the Stratonovich interpretation of the stochastic equation (\ref{eq3}),
the stationary solution of the corresponding Fokker-Planck equation can be
written as follows \cite{GO2001}:
\begin{equation}
\mathcal{P}_{st}[x]\propto
\exp\left(-\frac{\mathcal{U}_{ef}[x]}{\sigma^2}\right),
\end{equation}
where the effective potential functional $U_{ef}[x]$ is of the form
\begin{equation}\label{Ueff}
\mathcal{U}_{ef}[x]=\mathcal{U}[x]-\frac{\sigma^2}{2}\int{\rm d}\mathbf{r}\ln
D_{ef}[x].
\end{equation}
It follows, that the stationary distribution functional $\mathcal{P}_{st}$ or
the effective functional $\mathcal{U}_{ef}$ are obtained exactly: the form of
initial functional $\mathcal{U}$ is supposed to be known, the second term in
$\mathcal{U}_{ef}$ can be calculated if needed. Let us note, if we assume that
$\mathcal{U}$ plays a role of an effective free energy functional, then
rewriting the integral in Eq.(\ref{Ueff}) as $\mathcal{S}_{ef}=\int{\rm
d}\mathbf{r}\ln (D_{ef}[x])^{-1}$, the expression (\ref{Ueff}) can be
transformed into the thermodynamic relation between free energy, internal
energy and entropy functionals:
$\mathcal{U}_{ef}=\mathcal{U}+\sigma^2\mathcal{S}_{ef}$. Therefore, according
to such relation the noise intensity $\sigma^2$ addresses to an effective
temperature of the bath, whereas $S_{ef}$ plays a role of the effective
entropy. Such situation is well known in the stochastic systems theory. It
appears when the multiplicative fluctuations corresponded to the internal
noise. The last one results to the entropy change that yields entropy driven
phase transitions \cite{GO2001,Haeften2004,LB2006,EPJ2008}. In this paper we
will not discuss above phase transitions, we will consider an ability of the
noise to sustain or induce formation of spatial structures.

Stationary noise sustained structures $x_s(\mathbf{r})$ correspond to the
extreme of the functional $\mathcal{U}_{ef}[x]$. To find an associated equation
for $x_s(\mathbf{r})$ we make the first variation of the effective functional
with respect to $x$ and equal it to zero, that is
\begin{equation}
\delta \mathcal{U}_{ef}[x]=-\int{\rm d}\mathbf{r}\delta
x\left\{[f[x]D_{ef}[x]+D_{ef}[x]\nabla\left(D_{ef}[x]\nabla\mu[x]\right)+\frac{\sigma^2}{2D_{ef}[x]}\frac{\partial
D_{ef}[x]}{\partial x}\right\}=0.
\end{equation}
As a result, the corresponding equation takes the form
\begin{equation}\label{ge}
\nabla\left(D_{ef}[x]\nabla\mu[x]\right)+f[x]+\frac{\sigma^2}{2D^2_{ef}[x]}\frac{\partial
D_{ef}[x]}{\partial x}=0.
\end{equation}
Homogeneous states are defined as solutions of the reduced equation
\begin{equation}\label{bif}
fD^2_{ef}=-\frac{\sigma^2}{2}\frac{\partial D_{ef}}{\partial x}.
\end{equation}
From the mathematical viewpoint solutions of Eq.(\ref{bif}) give extreme
positions of the function $U_{ef}(x)$ obtained under supposition that
$x(\mathbf{r})=const$.

\section{The model}

In order to investigate a possibility of patterns formation we use the well
known assumptions for the diffusion coefficient $D_{ef}$: the diffusion
coefficient decreases from the constant value with the field $x$ growth.
Following Ref.\cite{Bray}, the generalized formula for $D_{ef}$ can be written
in the form $D_{ef}(x)=(1-x^2)^\alpha$, where $\alpha>0$. Under such an
assumption the diffusion $D_{ef}$ decreases slightly in the vicinity of $x=0$
and goes to zero with an increase in $x$ by absolute value. At small $\alpha$
one can use an approximation
\begin{equation}\label{Df}
D_{ef}=\frac{1}{1+\alpha x^2}, \quad \alpha\ge0.
\end{equation}
Such a construction assumes that fluctuations can not disappear with an
increase in $x$, moreover some analytic solutions of the problem can be
obtained.

The reaction term $f(x)$ can be defined according to a chemical kinetics and
generally is represented through a potential function, $V(x)$ in the standard
way: $f(x)=-\partial V/\partial x$. In this work we study a case of nonlinear
potential force of the form $f(x)=-\prod_i(x-x^0_{(i)})$, where the set
$\{x^0_{(i)}\}$ corresponds to zero values of the force and relates to
stationary points of the deterministic system. In our approach this force
associates with the potential
\begin{equation}\label{V(x)nsym}
V(x)=\frac{x^4}{4}+\frac{\mu}{3}x^3-\frac{\varepsilon}{2}x^2,
\end{equation}
here $\mu$ and $\varepsilon$ are constants that control the system dynamics.
The potential (\ref{V(x)nsym}) has three extrema placed at
$x_0^{\pm}=-\mu/2\pm\sqrt{\mu^2+4\varepsilon}/2$, and at $x_0=0$. A spinodal is
given by equation $\varepsilon=-\mu^2/4$. A prototype model of chemical
reactions is $A\rightleftarrows B$ with transient reactions $A+2X
\rightleftarrows 3X$, $X\rightleftarrows B$, where first reaction occur with
constant rates $k_1$, $k_2$, the second one is realized with $k_3$, $k_4$
\cite{schlogl}. Here $x\in[-1,1]$ measures deviations of spices $X$
concentration from the constant fixed value controlled by parameters
$\varepsilon$ and $\mu$, which relate to the constant rates $k_1$, $k_2$,
$k_3$, $k_4$.

To make the quantitative analysis let us use the chemical potential in the
standard form $\mu=\delta{\mathcal{F}}/\delta x$, where the free energy
functional is assumed in the Ginzburg-Landau form
\begin{equation}
\mathcal{F}=\int{\rm d}\mathbf{r}\left[\phi(x)+\frac{\beta}{2}(\nabla
x)^2\right],
\end{equation}
$\phi(x))$ the local potential; $\beta>0$ is an inhomogeneity constant. In oder
to make a general description of patterns formation scenario let us consider
the prototype model of phase separation where the local potential is of the
form
\begin{equation}\label{x4}
\phi=-\frac{\kappa}{2}x^2+\frac{1}{4}x^4,
\end{equation}
where $\kappa$ is a control parameter. From the formal view point one can
consider two different models. Indeed, in the simplest case of lateral
interactions one can use the local potential in the form
$\phi(x)=-\frac{\kappa}{2}x^2$, $\kappa>0$ \cite{ChemReact}. A case of ordinary
phase separation scenario corresponds to the $x^4$-construction, given by
Eq.(\ref{x4}). Next, we derive a formalism for a general model and compare
results for these two models of phase separation.

\section{Results}

\subsection{Noise induced transitions}

Let us note that the noise influence leads to a short-time instability in our
model. Indeed, the linear stability analysis of the linearized Langevin
equation (\ref{eq3}) allows to find an evolution equation for the structure
function $S(\mathbf{k},t)=\langle x_{\mathbf{k}}(t)x_{-\mathbf{k}}(t)\rangle$,
where $x(\mathbf{k},t)=\int x(\mathbf{r},t)e^{{\rm i}\mathbf{k}\mathbf{r}}{\rm
d}\mathbf{r}$ is the Fourier transform of the concentration field,
$\langle\ldots\rangle$ represents an ensemble average over noise. Considering
the state $x=0$, a spherically averaged structure function $S(k,t)=\int
S(\mathbf{k},t) {\rm d}{\Omega_k}$ (${\Omega_k}$ is the hyperspherical shell of
radius $k$) evolves according to a linear equation
\begin{equation}
\frac{1}{2}\frac{{\rm d}S(k,t)}{{\rm d}t}=-\omega(k)S(k,t)+\sigma^2,
\end{equation}
where $\omega(k)=k^2(\beta k^2-\kappa)-\varepsilon-\alpha\sigma^2$. It follows
that internal multiplicative noise leads to instability of the null state. The
stationary value of the structure function is
\begin{equation}
S_{st}(k)=\frac{\sigma^2}{k^2(\beta k^2-\kappa)-\varepsilon-\alpha\sigma^2}.
\end{equation}
It is seen that at short-time scales with an increase in the noise intensity
the peak position of the function $S_{st}(k)$ is shifted toward large values of
the wave vector.

To make an appropriate analysis of patterns formation scenario, let us consider
a case of a homogeneous (zero-dimensional) system, assuming stochastic variable
depending on the time only, i.e. $x(\mathbf{r},t)=x(t)$. The main attention
will be paid to stationary states investigation. In the case under
consideration the stationary probability density function is of the form
$P_{st}\propto\exp(-U_{ef}(x)/\sigma^2)$, where the functional is reduced to
the function, $\mathcal{U}_{ef}[x]\to U_{ef}(x)$. Hence, the effective
potential is $U_{ef}(x)=-\int {\rm d}x'f(x)D_{ef}(x)-(\sigma^2/2)\ln
D_{ef}(x)$. Inserting definitions for $f$ and $D_{ef}$ into this construction,
we obtain
\begin{equation}\label{Uef}
U_{ef}(x)=\frac{1}{2\alpha}x^2+\frac{\mu}{\alpha}x-\frac{1}{2\alpha}\left(\varepsilon+\frac{1}{\alpha}\right)\ln(1+\alpha
x^2 )
 -\frac{\mu}{\alpha^{3/2}}{\arctan}(\alpha\sqrt{x})+\frac{\sigma^2}{2}\ln(1+\alpha
 x^2).
\end{equation}
To consider a case $x\in[-1,1]$ let us assume values of both $\varepsilon$ and
$\mu$ to locate a minimum $U_{ef}(x_-)$ at $x<0$, a minimum $U_{ef}(x_+)$ we
locate at $x>0$. An appropriate choice is $\varepsilon=0.2$, $\mu=-0.5$,
$\alpha>0$; the corresponding dependence $U_{ef}(x,\sigma^2)$ is shown in
Fig.1.
\begin{figure}
\centering \includegraphics[width=90mm
]{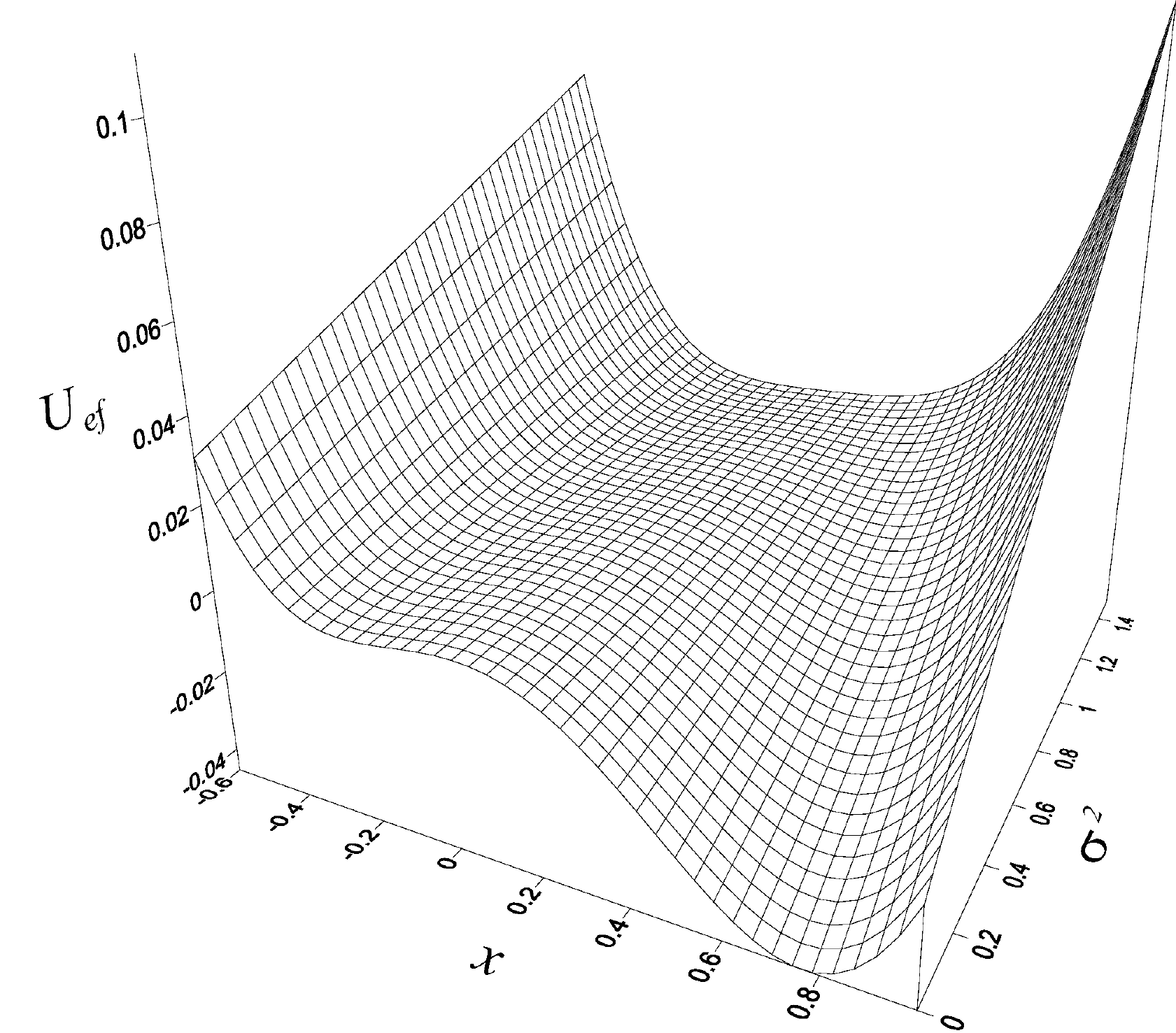} \caption{The effective potential $U_{ef}(x,\sigma^2)$ at
$\alpha=0.2$, $\varepsilon=0.2$, $\mu=-0.5$.\label{U}}
\end{figure}
To find most probable states we solve the problem ${\rm d}U_{ef}(x)/{\rm
d}x=0$. The corresponding equation takes the form (\ref{bif}). As it follows
from our consideration the functional dependence of $D_{ef}(x)$ leads to
bifurcations (a number of extrema of the effective potential changing). Hence,
solutions of Eq.(\ref{bif}) allow to find bifurcation diagram illustrating
extrema positions of the effective potential. In order to calculate critical
points related to the bifurcations one needs to differentiate Eq.(\ref{bif})
with respect to $x$ and after solve both obtained equation and Eq.(\ref{bif})
simultaneously. Solutions of Eq.(\ref{bif}) show that a root $x_0=0$ exists
always. Another two roots
\begin{equation}\label{xpm}
x_{\pm}=-\frac{\mu}{2}\pm\frac{1}{2}\sqrt{\mu^2+4\varepsilon-4\sigma^2\alpha}
\end{equation}
are realized if $\sigma^2<\sigma^2_c$ where
\begin{equation}
\sigma^2_c=\frac{1}{\alpha}\left(\varepsilon+\frac{\mu^2}{4}\right).
\end{equation}
At $\sigma^2=\sigma^2_c$ solutions $x_-$ and $x_+$ degenerate, and at
$\sigma^2>\sigma^2_c$ only trivial one, $x_0=0$, remains. The corresponding
dependencies $x_\pm(\sigma^2)$ are shown in Fig.2a. To understand
transformations of the system states let us use the noise induced transitions
formalism \cite{Horst}. As it follows from naive consideration, the bimodal
stationary distribution $P_{st}(x)\propto \exp(-U_{ef}(x)/\sigma^2)$ becomes
unimodal with an increase in the noise intensity $\sigma^2$. In the case under
consideration above transition occurs in the following manner. At $\sigma^2=0$
a form of the effective potential $U_{ef}$ is identical topologically to a form
of the initial potential $V(x)$. With an increase in the noise intensity
$\sigma^2$ a minimum of $U_{ef}(x_-)$ located at $x_-$ tends to zero, at
$\sigma^2=\sigma^2_s=\varepsilon/\alpha$ the effective potential has a double
degenerated point, $x_0=x_-=0$. Therefore, the values $\sigma^2_s$ define a
spinodal curve. At $\sigma_s^2<\sigma^2<\sigma^2_{0}$ the point $x_0$ relates
to a minimum, whereas $x_-$ defines a maximum position of the function
$U_{ef}$. These two minima differ in depth, i.e. $U_{ef}(0)>U_{ef}(x_+)$. At
$\sigma^2=\sigma^2_{0}$ one has $U_{ef}(0)=U_{ef}(x_+)$, therefore,
$\sigma^2_{0}$ defines a coexistence line (binodal). With a further increase in
$\sigma^2$ we get $U_{ef}(0)<U_{ef}(x_+)$. An equality $U_{ef}(x_-)=
U_{ef}(x_+)$ is satisfied at $\sigma^2=\sigma^2_{c}$, hence the bifurcation
point, $\sigma^2_{c}$, define another spinodal. At $\sigma^2>\sigma^2_c$ the
effective potential has one well only. Therefore, in such a noise induced
transition we have shift of the potential extreme, transformation of the global
minimum into a local one, loss of its stability and, finally, change a number
of extreme of the function $U_{ef}$. The corresponding diagram illustrating
above situation is shown in Fig.2b.
\begin{figure}
\centering a \hspace{70mm} b\\
\includegraphics[width=60mm
]{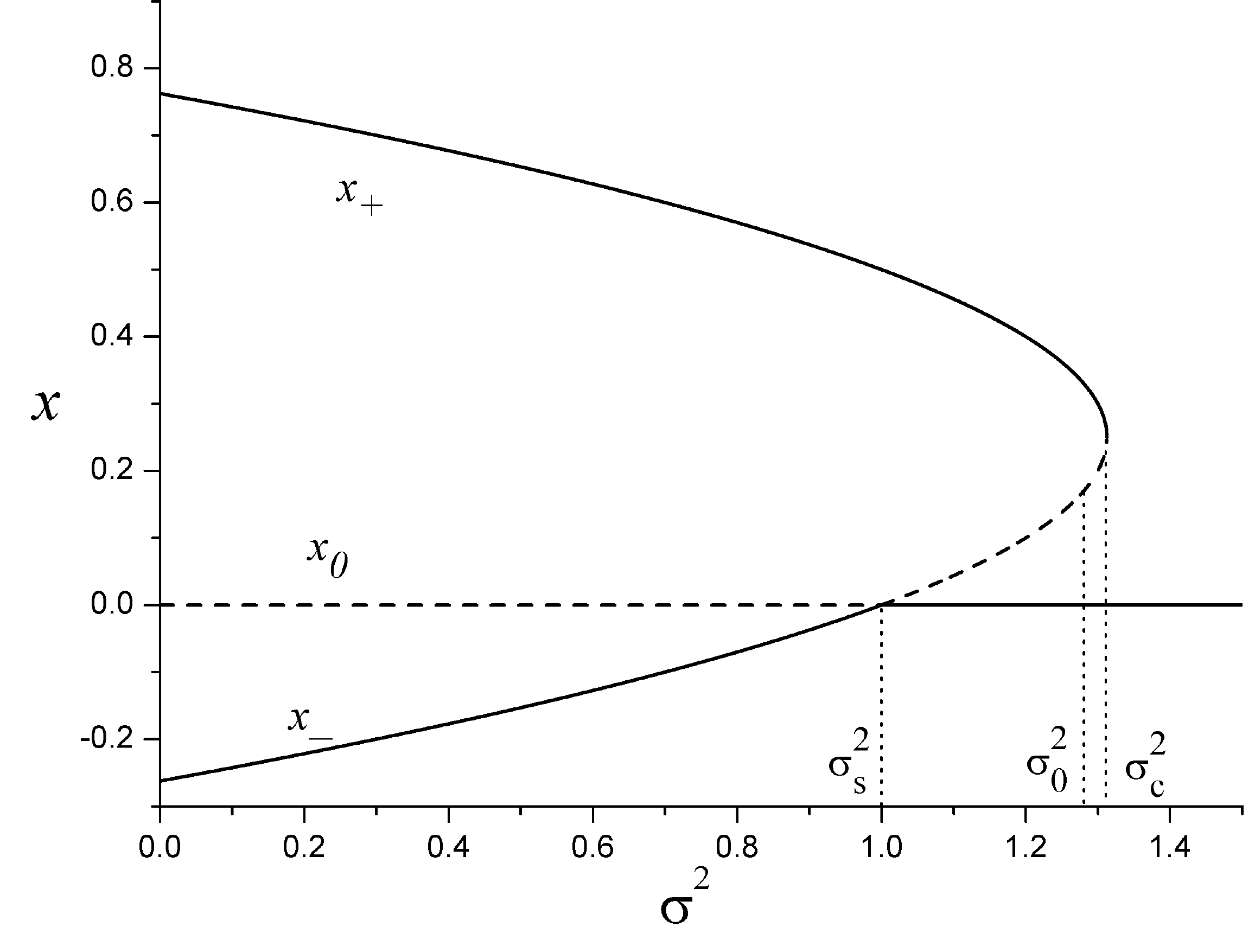}
\includegraphics[width=60mm
]{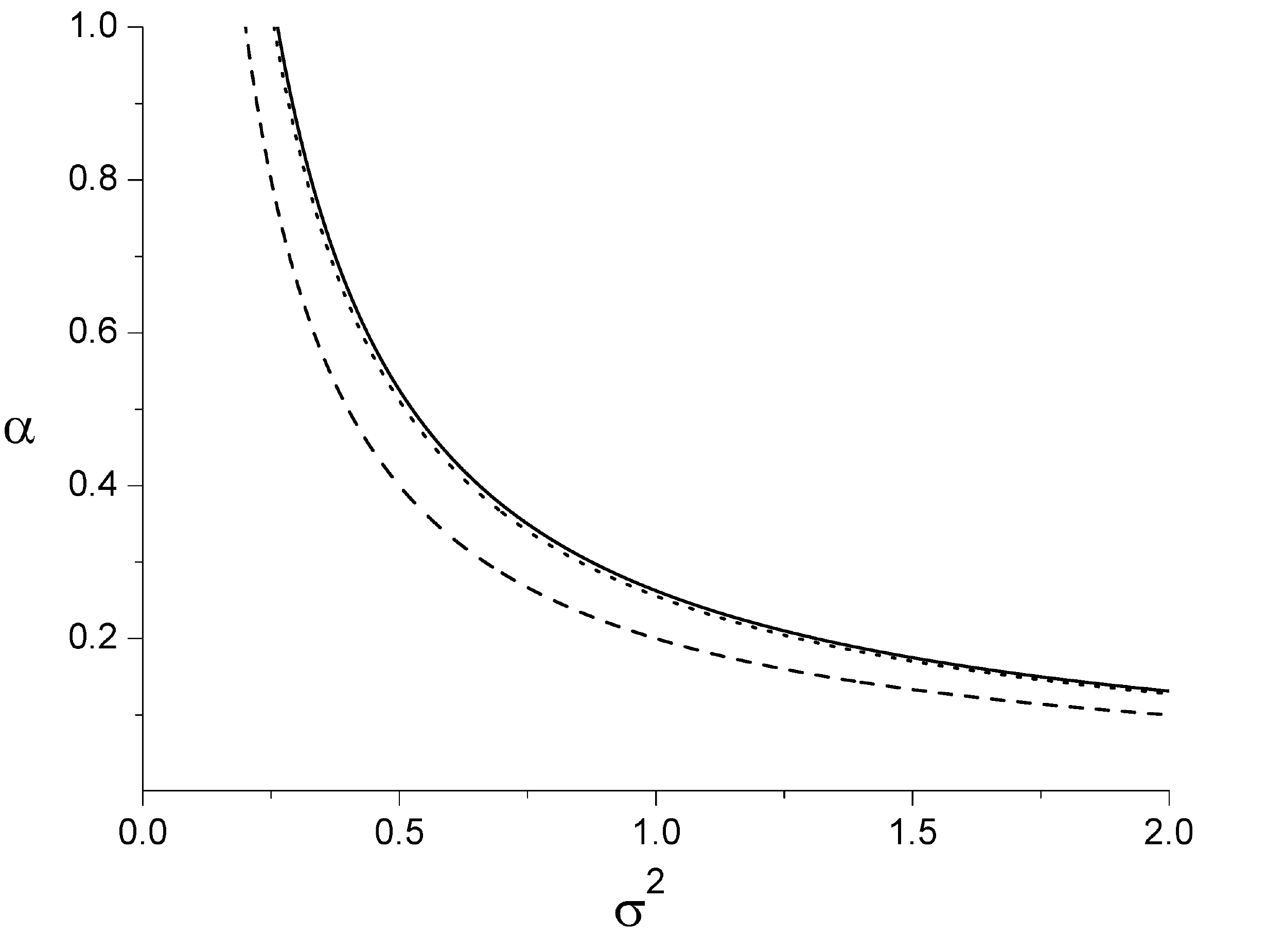}
 \caption{Bifurcation diagram (a) and phase diagram (b) for noise
induced transitions (a change number of extrema of the function $U_{ef}(x)$) at
$\alpha=0.2$, $\varepsilon=0.2$, $\mu=-0.5$. In plot (a): solid lines define
stable states, dashed line corresponds to unstable solution. In plot (b): solid
line corresponds to $\sigma^2_c$ values, dashed and dotted lines relate to
$\sigma_s^2$ and $\sigma^2_0$, respectively. \label{NIT}}
\end{figure}

\subsection{Noise induced patterns}

Next, let us consider a possibility of the system to manifest stationary
structures formation. To this end we assume the spatial dependence of the
stochastic field and solve the variation problem, $\delta
\mathcal{U}_{ef}[x]/\delta x=0$. The corresponding equation takes the form of
Eq.(\ref{ge}). Let us rewrite Eq.(\ref{ge}) in a more convenient form
\begin{equation}\label{st_lat}
D_{ef}\left[\beta\nabla^4x-\frac{\partial^2 \phi}{\partial x^2}\nabla^2
x-\frac{\partial^3 \phi}{\partial x^3}(\nabla x)^2\right]-\frac{\partial
D_{ef}}{\partial x}\frac{\partial^2 \phi}{\partial x^2} (\nabla
x)^2=f+\frac{\sigma^2}{2D^2_{ef}}\frac{\partial D_{ef}}{\partial x}.
\end{equation}
As it follows from our consideration, the system stationary behaviour can be
described in the 4-dimensional space $(x,y,z,u)$ where $y=\nabla x$, $z=
\nabla^2x$, $u=\nabla^3 x$. Using results form the homogeneous system analysis
one can state that there are three fixed points describing homogeneous states
with coordinates $(x_-,0,0,0)$, $(0,0,0,0)$ and $(x_+,0,0,0)$ at
$\sigma^2<\sigma^2_c$ and the unique fixed point $(0,0,0,0)$ at
$\sigma^2>\sigma^2_c$.

Firstly, let us investigate the stability of stationary solutions of
Eq.(\ref{st_lat}) in the $\mathbf{r}$--space. To this end one assumes
$x(\mathbf{r})$ to be in the form
$x(\mathbf{r})=x(\mathbf{0})+{\delta}\exp(\vec\lambda\mathbf{r})$, $\delta\ll
1$ is a small perturbation in the vicinity of the corresponding homogeneous
solution. Inserting this assumption with corresponding expressions for all
possible derivatives into Eq.(\ref{st_lat}), one gets the equation
\begin{equation}\label{l4}
\lambda^4-\lambda^2\left.\left[D_{ef}\frac{\partial^2 \phi}{\partial
x^2}\right]\right|_{x=\{x_0,x_-,x_+\}}-\left.\left[\frac{\partial }{\partial
x}\left(f+\frac{\sigma^2}{2D^2_{ef}}\frac{\partial D_{ef}}{\partial
x}\right)\right]\right|_{x=\{x_0,x_-,x_+\}}=0
\end{equation}
for eigenvalues $\vec \lambda=\vec\gamma\pm {\rm i}\vec k$ of the Jacobi
matrix, where $\vec\gamma=\{\gamma_j\}_{j=1}^4$, $\vec k=\{k_j\}_{j=1}^4$;
$\gamma_j=\Re{\lambda_j}$ defines the local stability of the fixed point along
$j$-th axis, $k_j=\Im{\lambda_j}$ gives the corresponding oscillations
frequency that relates to the wave vector magnitude.

From the linear stability analysis it follows that relations for imaginary
parts $k_1=-k_2$, $k_3=-k_4$ and real parts $\gamma_1=-\gamma_2$,
$\gamma_3=-\gamma_4$ are hold. Solving Eq.(\ref{l4}) and using these
symmetrical properties, we present only positive values of $\vec k$ and
$\vec\gamma$ as functions of the noise intensity $\sigma^2$ in Fig.3.
\begin{figure}[!h]
\centering
 a)\includegraphics[
 width=60mm]{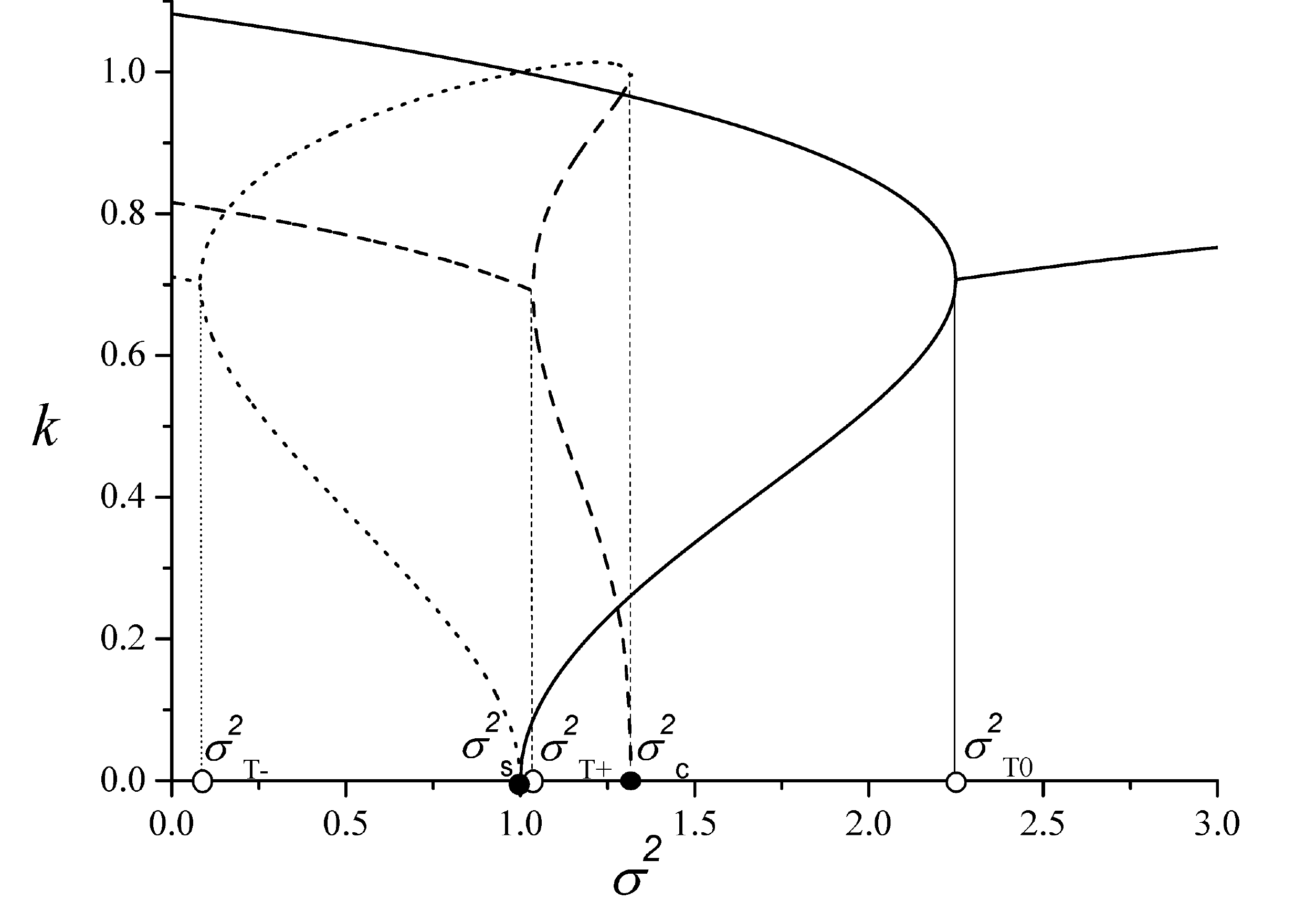} \ \includegraphics[
 width=60mm]{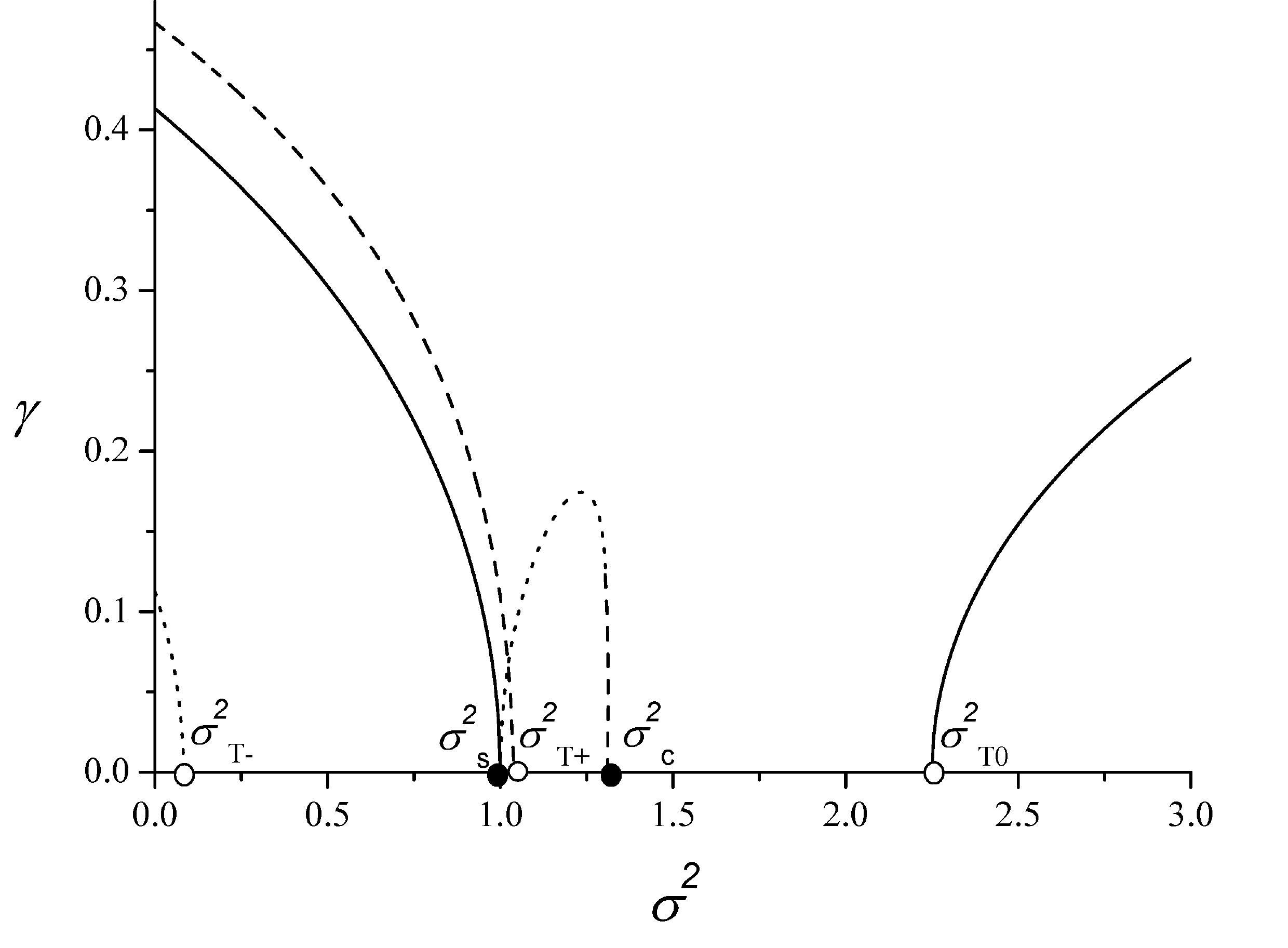}
 b)\includegraphics[
 width=60mm]{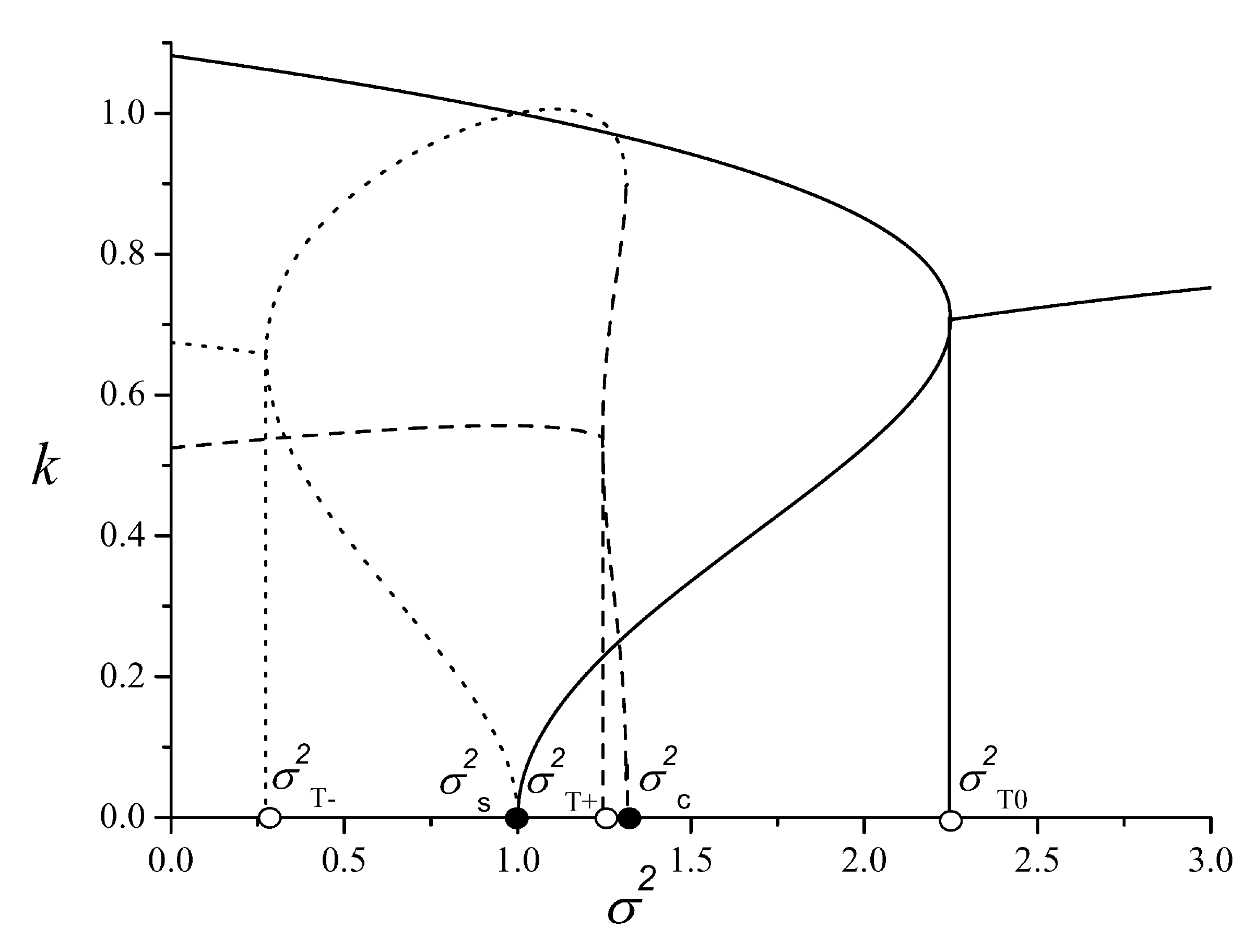} \ \includegraphics[
 width=60mm]{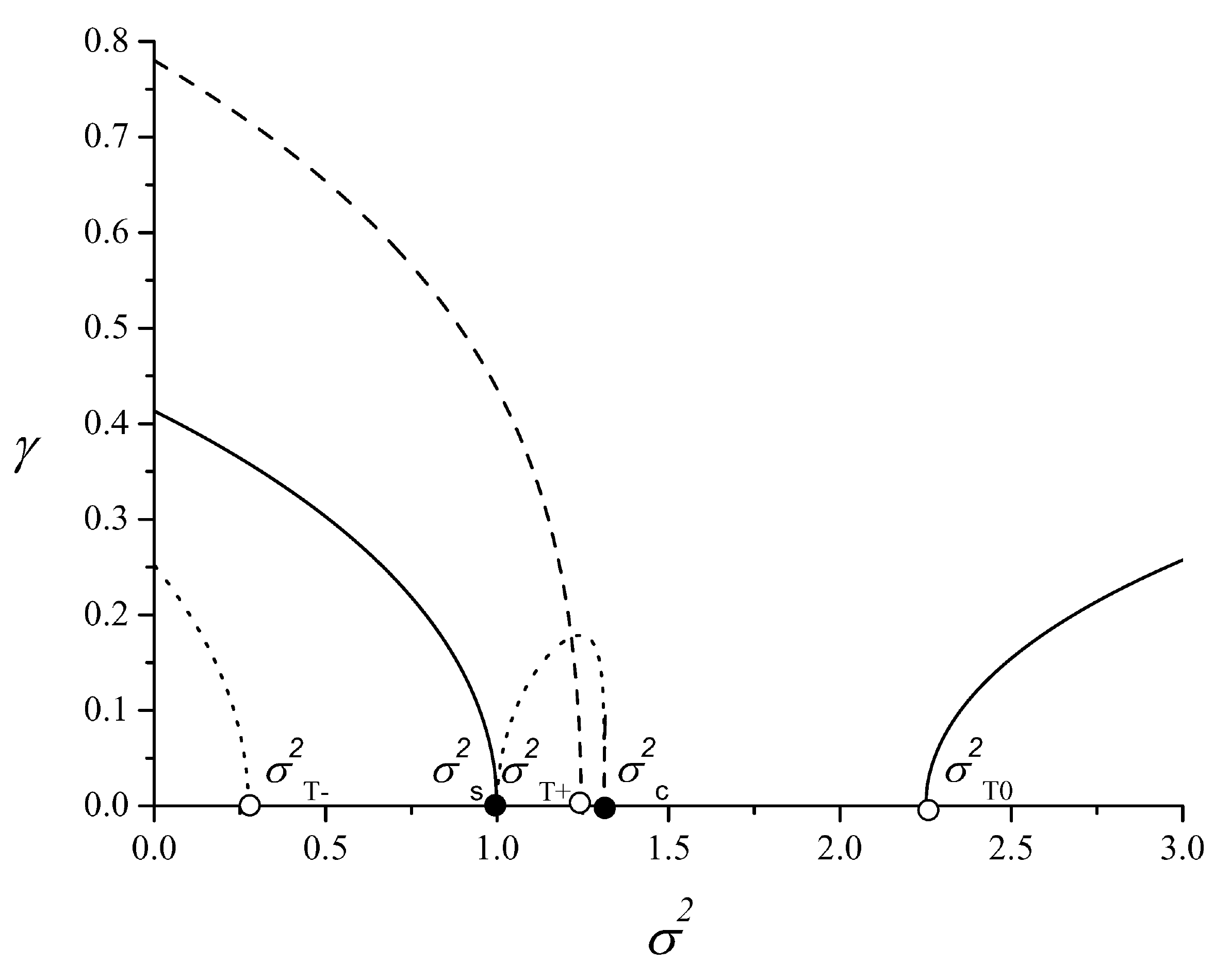}
 \caption{Positive
values of imaginary $\vec k$ (a) and real $\vec \gamma$ (b) parts of
eigenvalues $\vec\lambda$ versus the noise intensity $\sigma^2$, other
parameters are: $\varepsilon=0.2$, $\mu=0.5$, $\alpha=0.2$, $\kappa=1$. Solid
dashed and dotted lines correspond to stability of the fixed points
$O(0,0,0,0)$, $(x_+,0,0,0)$ $(x_-,0,0,0)$, respectively. Plots (a) correspond
to lateral interaction model, plots (b) are related to
$x^4$-model.\label{k_g(s)}}
\end{figure}
Obviously, if one of the real part related to the fixed point of a homogeneous
state is positive, then the such point is unstable in the $\mathbf{r}$--space.
In the case of zero values of all real parts for the fixed point with nonzero
imaginary ones, the fixed point is a center of a manifold type of limit cycle
(with a single nontrivial $k_j$) or torus (with more than one nonzero imaginary
parts). To relate $k_j$ to the profile $x(\mathbf{r})$ or the corresponding
derivatives we have solved Eq.(\ref{st_lat}) in the one-dimensional case at the
appropriate values for the system parameters. The corresponding description
will be given below. As Fig.3 shows at small noise intensities with
$\sigma^2<\sigma^2_{T-}$, all fixed points are unstable, no patterns can be
formed. If the threshold $\sigma^2_{T-}$ is crossed, then the stability of the
point $(x_-,0,0,0)$ is changed. It becomes a center of a set of tori with two
frequencies $k_1$ and $k_2$ displayed as dotted lines in Fig.3. When the
threshold $\sigma^2_s$ is crossed, then the stability of the solution
$(x_-,0,0,0)$ is changed again, it becomes an unstable one. Therefore, in the
interval $\sigma^2_{T-}\le\sigma^2\le\sigma^2_{s}$ the phase space is
characteized by a single set of tori formed at the center of the point
$(x_-,0,0,0)$. At $\sigma^2>\sigma^2_s$ the null state, characterized by the
point $(0,0,0,0)$ acquires a neutral stability, which is realized till
$\sigma^2=\sigma^2_{T0}$ (see solid lines in Fig.3), where
\begin{equation}
\sigma^2_{T0}=\sigma_s^2+\frac{\kappa^2}{4\alpha}.
\end{equation}
Moreover, at $\sigma^2=\sigma^2_{T+}$ the fixed point $(x_+,0,0,0)$ becomes a
center of an additional set of tori (dashed lines). Hence, in the interval of
the noise intensity $\sigma^2_{T+}<\sigma^2<\sigma^2_{c}$ two set of tori are
observed. At $\sigma^2=\sigma^2_c$ the stability of the point $(x_+,0,0,0)$ is
changed, it becomes unstable due to the one of the corresponding real part is
positive. Therefore, two sets of tori around the points $O(0,0,0,0)$ and
$(x_+,0,0,0)$ are realized till the threshold $\sigma^2_c$. Due to the
bifurcation related to the noise induced transition oscillations in the
vicinity of the point $(x_+,0,0,0)$ disappears at $\sigma^2=\sigma_c^2$ and as
a result a single set of tori around the point $(0,0,0,0)$ can be observed.

The value $\sigma^2_{T0}$ as other thresholds except $\sigma^2_s$ and
$\sigma_c^2$ are dependent on the parameter $\kappa$. Above, we have considered
the case when $\sigma_c^2<\sigma^2_{T0}$. From our analysis it follows that at
small noise intensities or without noise the homogeneous states are absolutely
unstable. With an increase in the noise intensity two-period solutions of
Eq.(\ref{st_lat}) are realized. At large noise intensity the system is
characterized by the unstable homogeneous solution. Therefore, we get a
reentrant picture of self-organization with the noise growth. The corresponding
phase diagram illustrating patterns formation is shown in Fig.4.
\begin{figure}[!h]
\centering
 a) \includegraphics[width=60mm
 ]{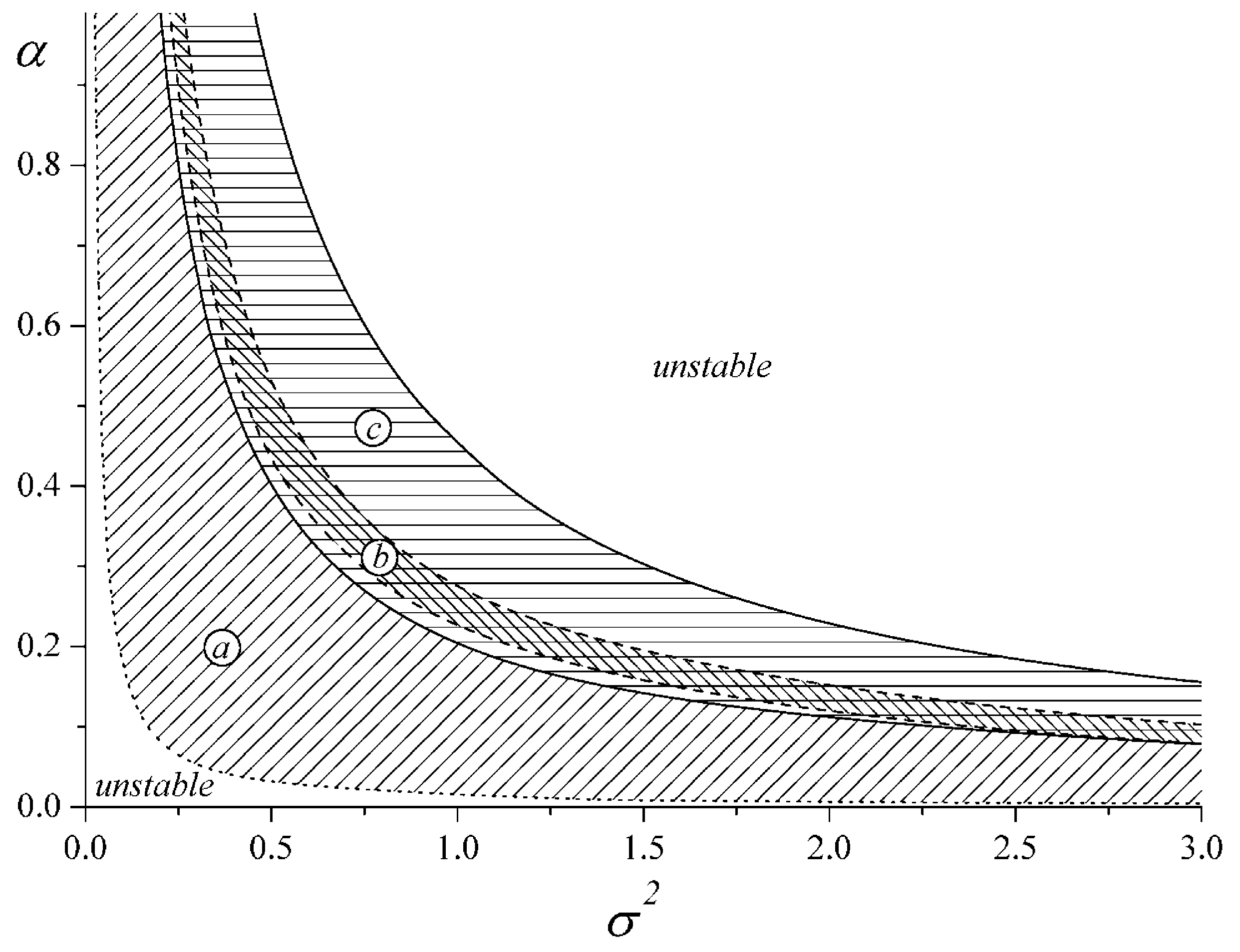} b) \includegraphics[width=60mm
 ]{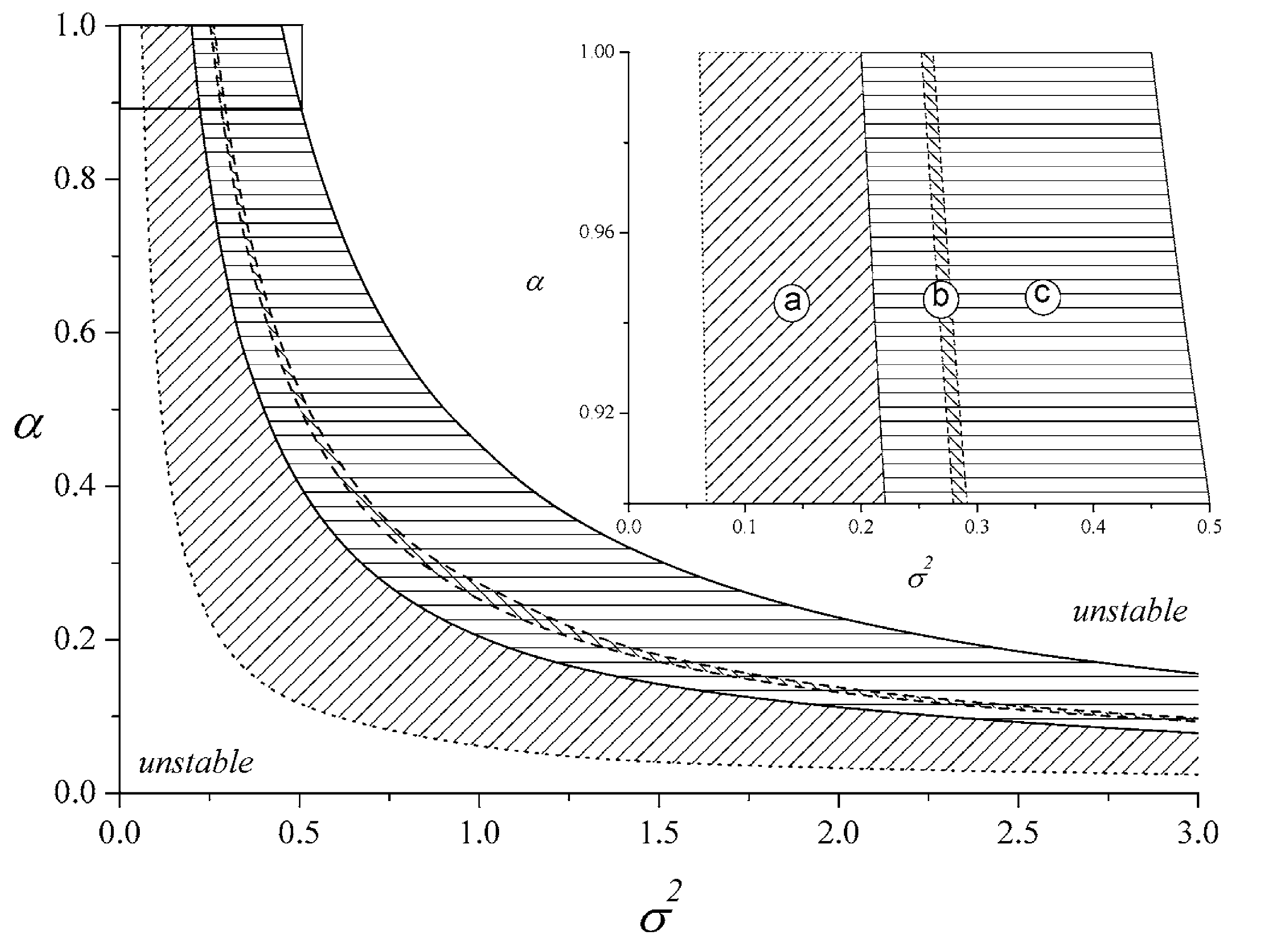}
 \caption{Phase
diagram of reentrant behaviour of self-organization process at
$\varepsilon=0.2$, $\mu=-0.5$, $\beta=1$, $\kappa=1.0$. Domains of tori
formation are filed: (a) corresponds to a tori set around the point
$(x_-,0,0,0)$; (c) is addressed to a tori set around the fixed point
$(0,0,0,0)$; double filed domain (b) relates to two set of tori around points
$(x_-,0,0,0)$ and $(x_+, 0,0,0)$. Plot (a) correspond to lateral interaction
model, (b) is related to $x^4$-model.\label{a(s)}}
\end{figure}
Here we plot the parameter $\alpha$ versus noise intensity $\sigma^2$ at fixed
values for other control parameters. Unstable states are outside the filled
domain. In the filled part we denote three sub-domains (a), (b) and (c). In the
sub-domains (a) and (c) only one set of tori is formed around points
$(x_-,0,0,0)$ and $(0,0,0,0)$, respectively. The domain (b) is characterized by
two sets of tori in the vicinity of both points $(0,0,0,0)$ and $(x_+,0,0,0)$.
All calculations are performed for two choices lateral interactions and
$x^4$-model of phase separation. Comparing Fig.3a and Fig.3b, one can see that
there are no qualitative changes in the dependencies of the eigenvalues of the
Jacobi matrix versus the noise intensity. The situation is similar for these
two models. The corresponding phase diagrams for two models of phase
separations are topologically identical. In the $x^4$-model the size of the
domain of patterns formation in the vicinity of the point $(x_+, 0, 0,0)$ is
reduced.

Let us relate every root $k_i$ to the corresponding solution of the stationary
equation (\ref{st_lat}), namely to $x(r)$, $y=\nabla x(r)$, $z=\nabla^2 x(r)$,
$u= \nabla^3 x(r)$. To make a more accurate analysis we use conditions that a
ratio of two different wave vectors $k_i/k_j$ gives a rational number and
appropriate real values equal zero, $\gamma_i=\gamma_j=0$, where $i\ne j$. We
plot all solutions in Fig.5 in a vicinity of every fixed point $(x_-,0,0,0)$,
$(x_+,0,0,0)$ and $(0,0,0,0)$ \footnote{To identify that the ratio $k_i/k_j$ is
about rational or irrational number we use an additional criteria. Due to in
numerical calculations all numbers are understood as rational ones with given
precision in our analysis we have used an algorithm allowing to determine a
fractal correlation dimension $D_c$ of a manifold (torus) in the space ($x$,
$\nabla x$, $\nabla^2 x$, $\nabla^3 x$). In our calculations the value
$D_c\simeq 1$ corresponds to the rational ratio $k_i/k_j$ (the solution of the
problem (\ref{st_lat}) is a three-dimensional closed line (torus) in the space
($x$, $\nabla x$, $\nabla^2 x$, $\nabla^3 x$)), whereas $D_c\simeq 2$ relates
to the irrational one (the corresponding solution is an unclosed line on the
torus in the four--dimensional space).}. It is seen that the profile $x(r)$ has
one large period related to small values of $k$. Two-period solutions are
related to profiles of spatial derivatives of the field $x$. Therefore, the
stationary noise sustained structures $x_{st}(r)$ has one period defined by
branches with small $k$ values in Fig.3. The corresponding solutions for
$x^4$-model are of the same kind, no qualitatively changes are observed.

\begin{figure}[!t]
\centering
 a)\includegraphics[width=40mm]{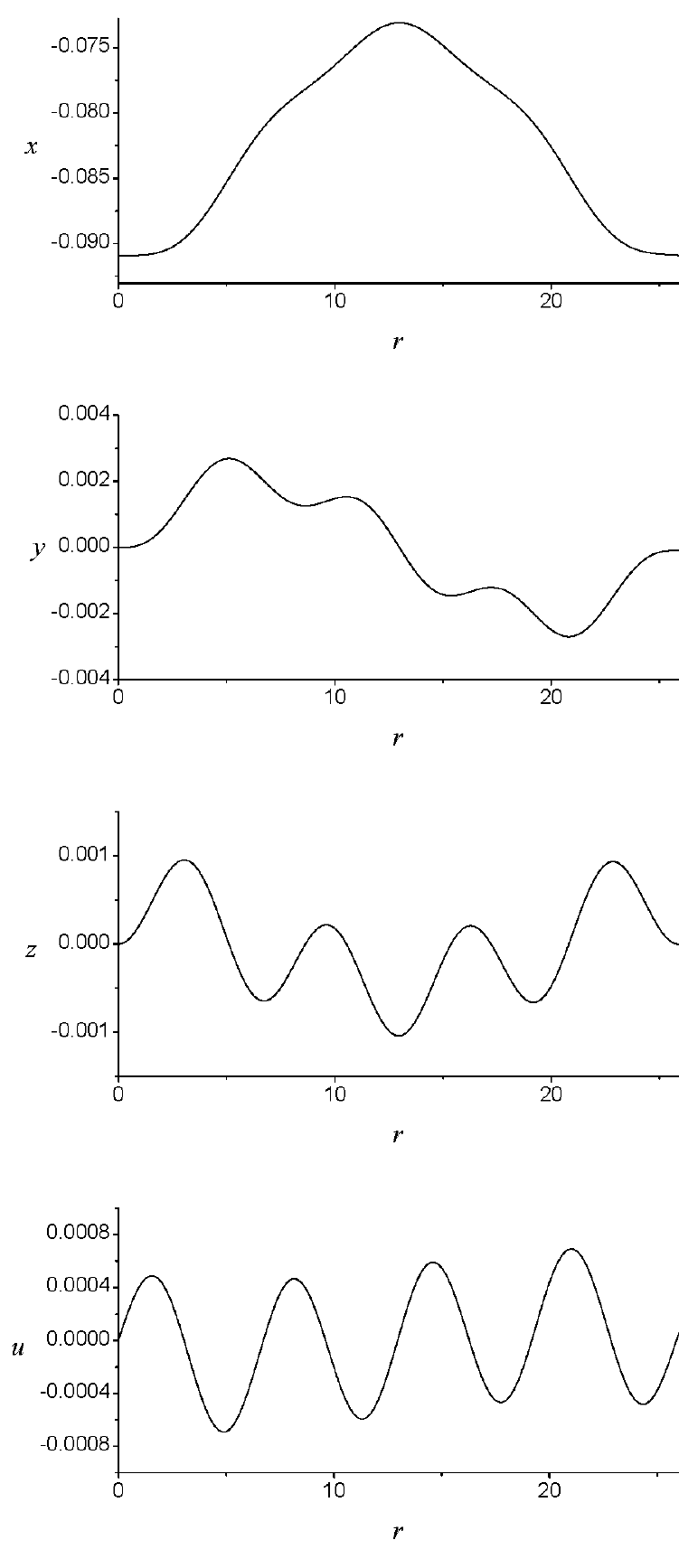}
 b)\includegraphics[width=40mm]{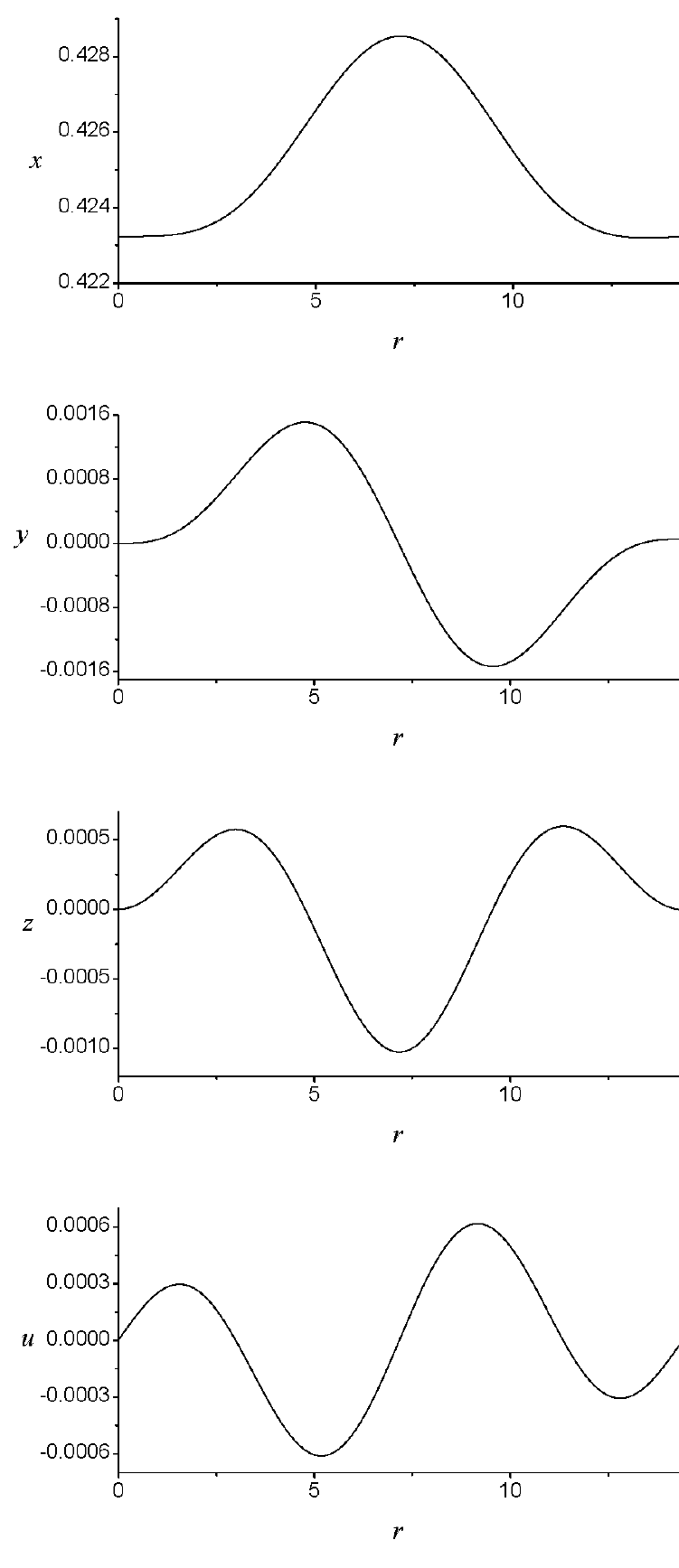}
 c)\includegraphics[width=40mm]{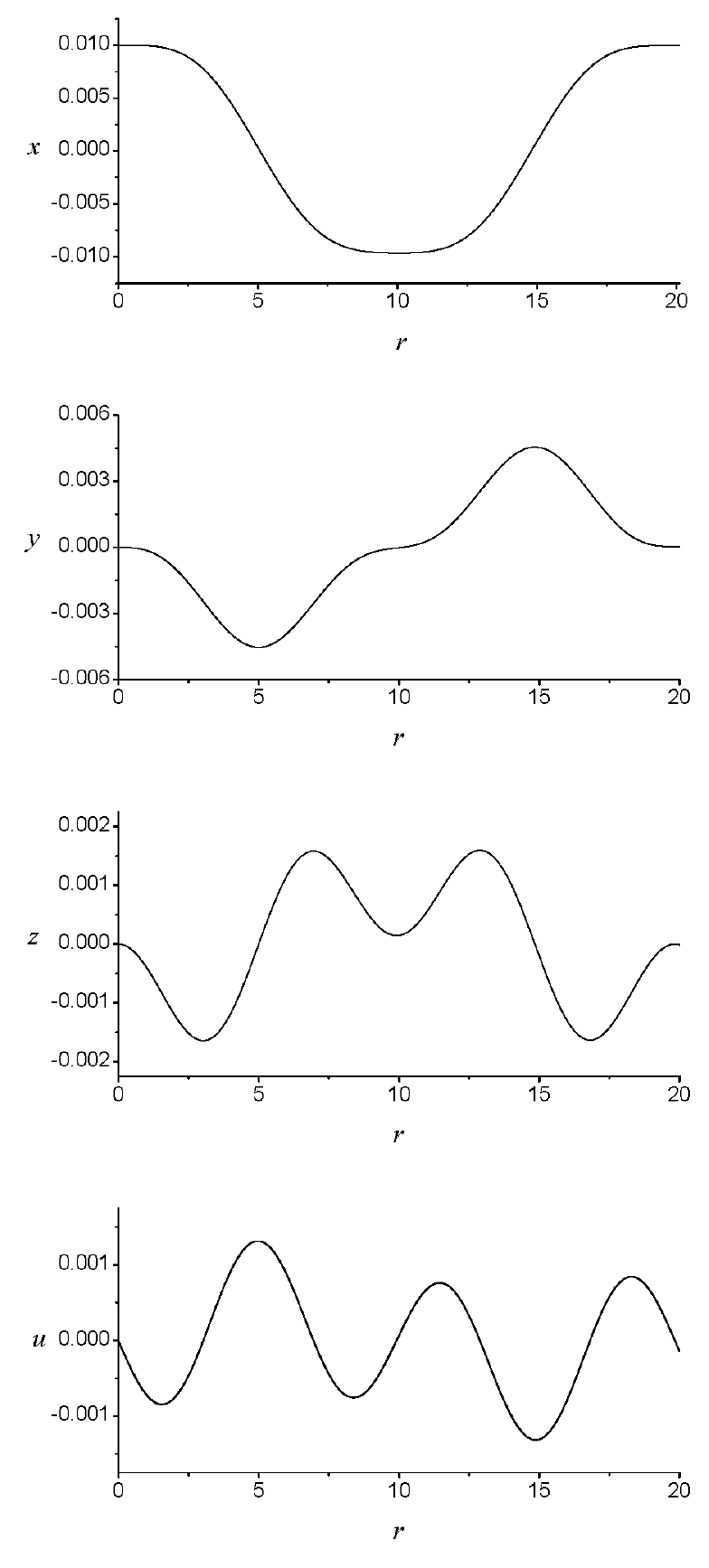}
 \caption{Solutions of the stationary
equation (\ref{st_lat}) spatial profiles for $x(r)$, $y(r)=\nabla x$,
$z(r)=\nabla y$, and $u(r)=\nabla z$: (a) $\sigma^2=0.75832$; , $x(0)=-0.083$,
$k_i/k_j=4$; b) $\sigma^2=1.15896$, $x(0)=0.425$, $k_i/k_j=2$; c)
$\sigma^2=1.45(45)$, $x(0)=10^{-5}$, $k_i/k_j=3$. Other parameters are:
$\alpha=0.2$, $\varepsilon=0.2$, $\mu=-0.5$, initial conditions for derivatives
are $y(0)=z(0)=u(0)=0$. \label{torus_x}}
\end{figure}

\begin{figure}[!t]
\centering
 a)\includegraphics[width=40mm]{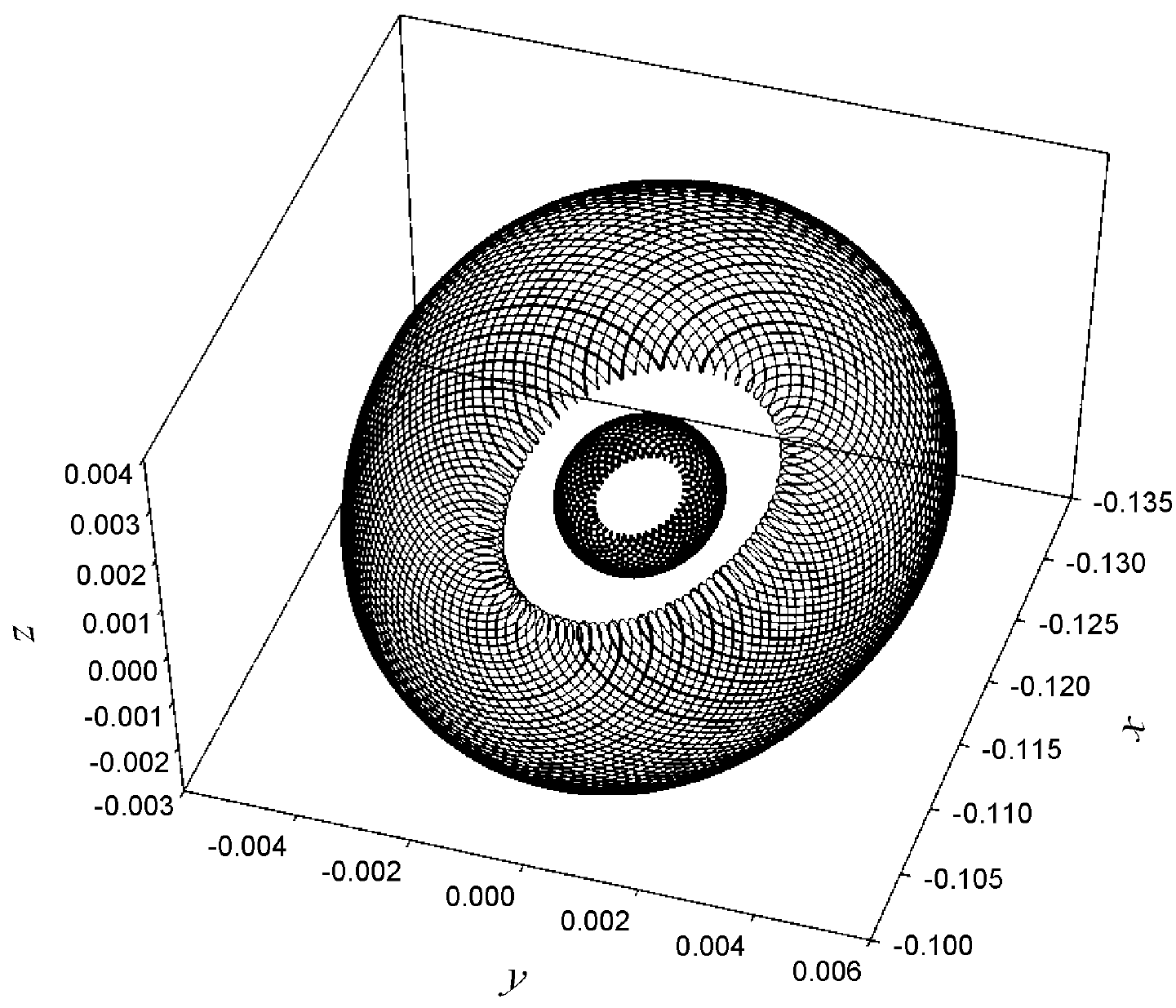}
 b)\includegraphics[width=40mm]{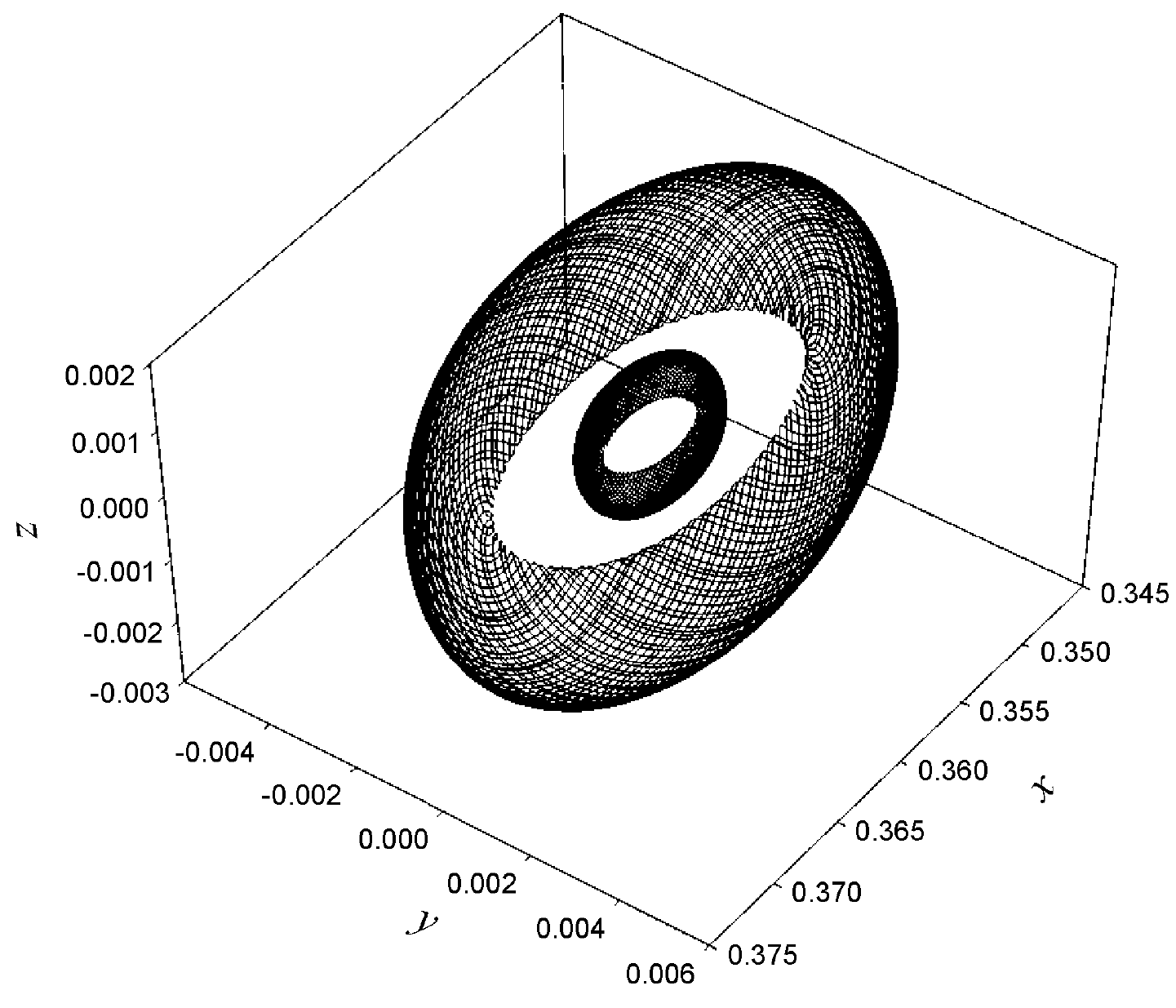}
 c)\includegraphics[width=40mm]{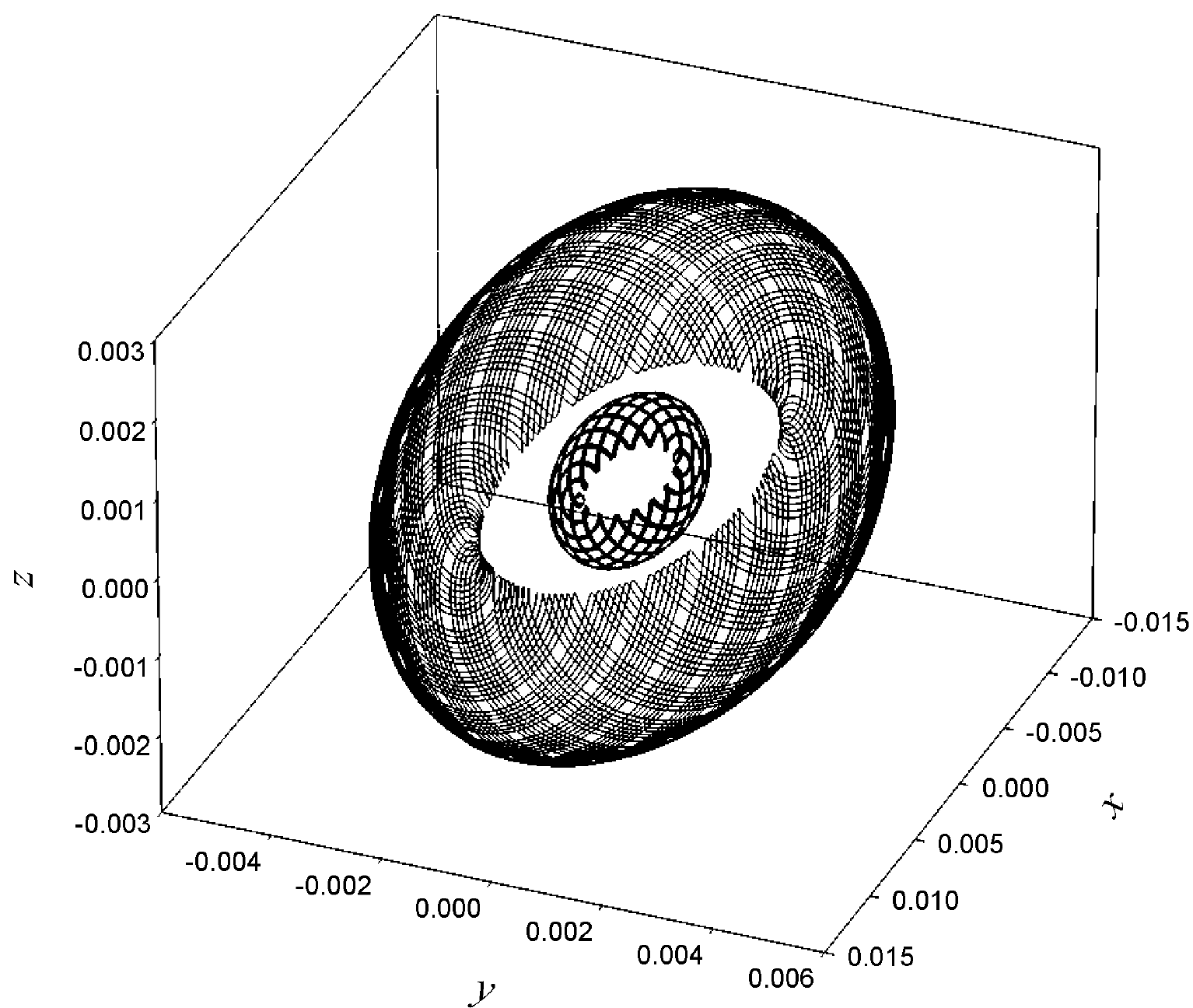}
 \caption{Solutions of the stationary
equation (\ref{st_lat}) (set of tori) in a space ($x$, $y=\nabla x$, $z=\nabla
y$) at ratio $k_i/k_j\simeq\pi$: a) around fixed point $x_-$ at
$\sigma^2=0.635213$; b) around fixed point $x_+$ at $\sigma^2=1.2532123$; c)
around fixed point $x_0$ at $\sigma^2=1.41767865$. Other parameters are:
$\alpha=0.2$, $\varepsilon=0.2$, $\mu=-0.5$, initial conditions for derivatives
are: $y(0)=z(0)=u(0)=0$. \label{torus_xyz}}
\end{figure}

Solutions of the problem (\ref{st_lat}) with irrational ratio $k_i/k_j$ are
shown in Fig.6. Here we choose the noise intensity values to be about the well
known irrational number with a good precision, $k_i/k_j\simeq\pi$. It follows
that in the three-dimensional functional space solutions of the problem
(\ref{st_lat}) is a three-dimensional line which lies on the torus. This line
is an unclosed and the corresponding torus is dense filled (the corresponding
fractal correlation dimension is $D_c\simeq 2.0\pm0.02$).

\subsection{Simulations}

To investigate spatial structures numerically let us consider solutions of the
Langevin equation (\ref{eq3}) numerically. In order to perform numerical
analysis, we redefine the model considering a regular two-dimensional lattice
with $N^2$ points and a lattice spacing $\ell$. Then, the partial differential
equation (\ref{eq3}) is reduced to a set of usual differential equations
written for an every cell $i$ on a grid in the form
\begin{equation}
\frac{{\rm
d}x_i}{\rm{d}t}=f(x_i)+(\nabla_L)_{ij}D(x_{j})(\nabla_R)_{jl}\frac{\partial
F}{\partial x_l}+\frac{1}{\sqrt{D(x_{i})}}\xi_i(t),
\end{equation}
where index $i$ labels cells, $i=1,\ldots, N^2$; the discrete left and right
operators are introduced
\begin{equation}
\begin{split} &(\nabla_L)_{ij}=\frac{1}{\ell}(\delta_{i,j}-\delta_{i-1,j}),\quad
(\nabla_R)_{ij}=\frac{1}{\ell}(\delta_{i+1,j}-\delta_{i,j}),\\
&(\nabla_L)_{ij}=-(\nabla_R)_{ji},\quad (\nabla_L)_{ij}
(\nabla_R)_{jl}=\Delta_{il}=\frac{1}{\ell^2}(\delta_{i,l+1}-2\delta_{i,l}+\delta_{i,l-1}),
\end{split}
\end{equation}
all sums are taken according to the Einstein rule. A derivative of the free
energy $F$ is of the form
\begin{equation}
\frac{\partial F}{\partial x_i}=\frac{{\rm d}\phi(x_i)}{{\rm
d}x_i}-\beta\Delta_{ij}x_j.
\end{equation} For stochastic sources the discrete correlator is of the form
$\langle\xi_i(t)\xi_j(t)\rangle=2\ell^2\sigma^2\delta_{ij}\delta(t-t')$. In the
following analysis we use the Stratonovich interpretation of the Langevin
equations (\ref{eq3}). In our integration procedure we choose $\ell=1$. After
discretization scheme described above, the resulting set of coupled stochastic
ordinary differential equations has been integrated numerically. To simulate
evolution of the system we have used a first-order Euler algorithm with a time
step $\delta t=0.005$. Initial conditions for the field were chosen from a
random Gaussian distribution with fixed average $\langle
x(\mathbf{r},0)\rangle$ (in vicinity of the fixed point defined from analysis
of the effective potential $U_{ef}(x)$, see Fig.2) and small dispersion,
$\langle(\delta x(\mathbf{r}),0)^2\rangle=0.003$. Typical patterns appeared in
a course of evolution at different values of the noise intensity $\sigma^2$ are
shown in Fig.7. Evolution of the two first moments is shown in Fig.8. From
numerical solutions it follows that at small noise intensity (see Fig.7a) the
system evolves by a scenario when inside a matrix phase a new phase appears.
\begin{figure}[!t]
\centering
 a)\includegraphics[width=40mm]{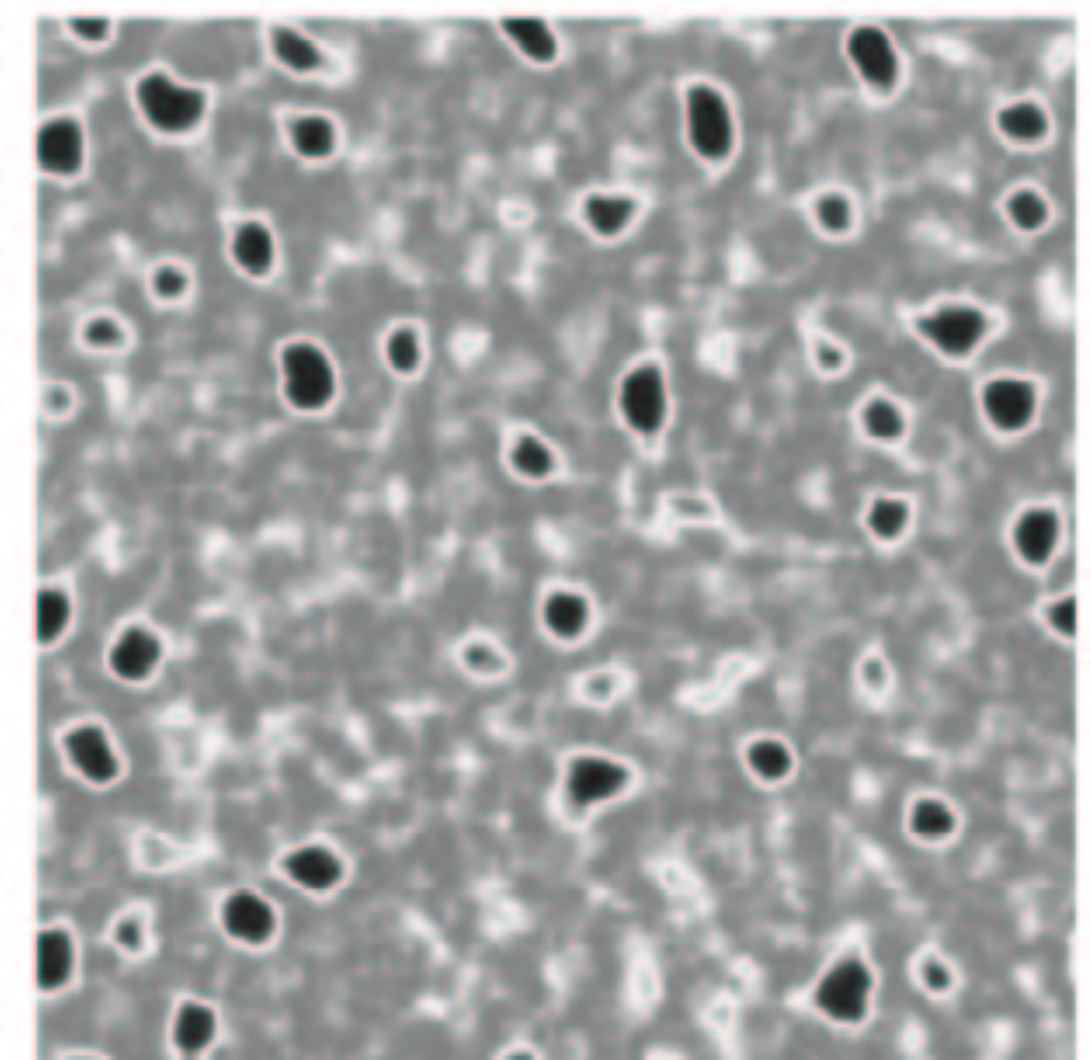}\ b)\includegraphics[width=39mm]{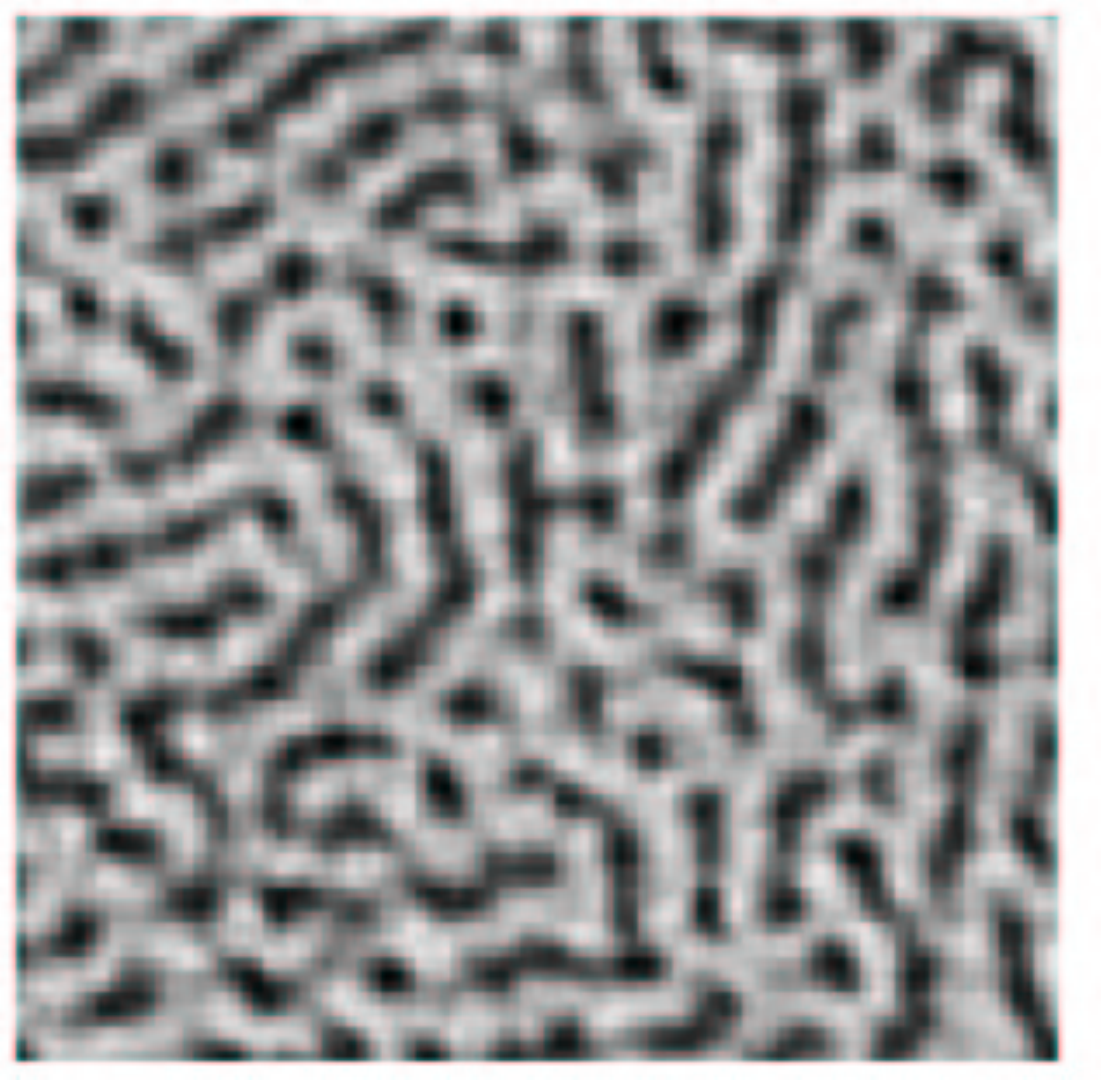} \ c)\includegraphics[width=38mm]{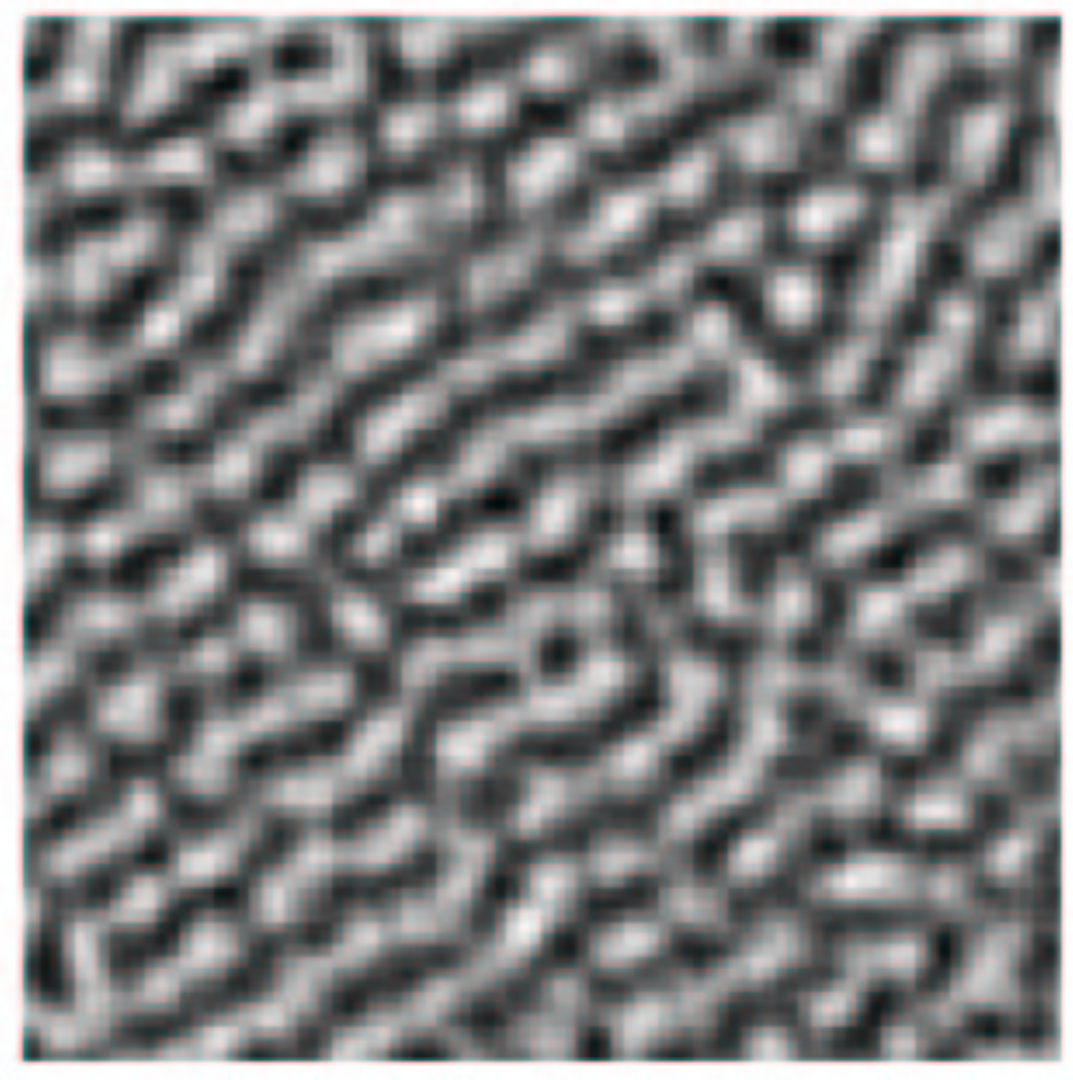}
 \caption{Spatial patterns
appeared in a course of evolution at $t=4000$: a) $\sigma^2=0.2$; b)
$\sigma^2=\sigma^2_0=1.2757$; c) $\sigma^2=2.0$. Other parameters are:
$\varepsilon=0.2$, $\mu=0.5$, $\kappa=1.0$, $\beta=1.0$,
$\alpha=0.2$\label{Stoch_pat}}
\end{figure}
\begin{figure}[!t]
\centering
 a)\includegraphics[
 width=40mm]{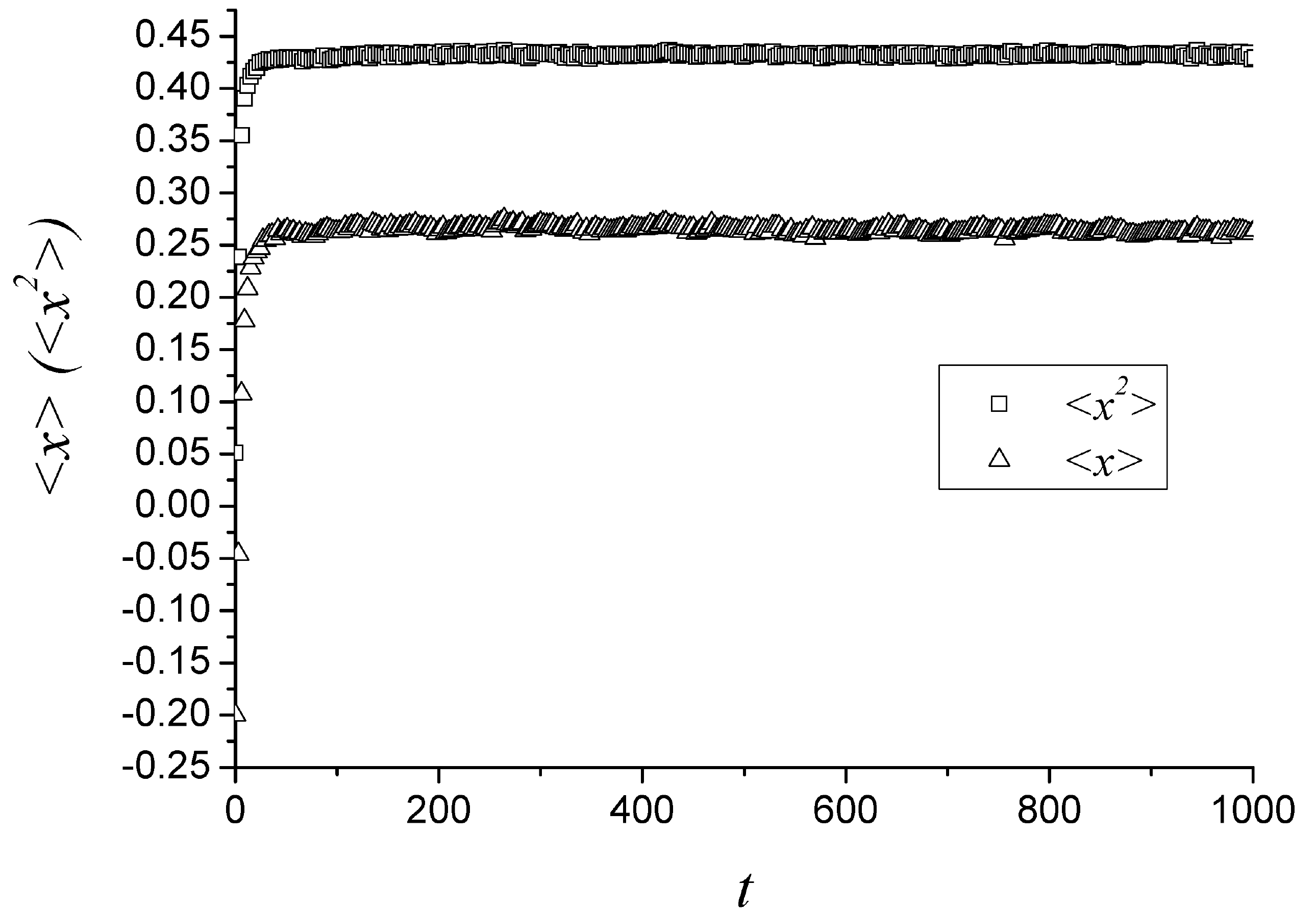}\ b)\includegraphics[
 width=40mm]{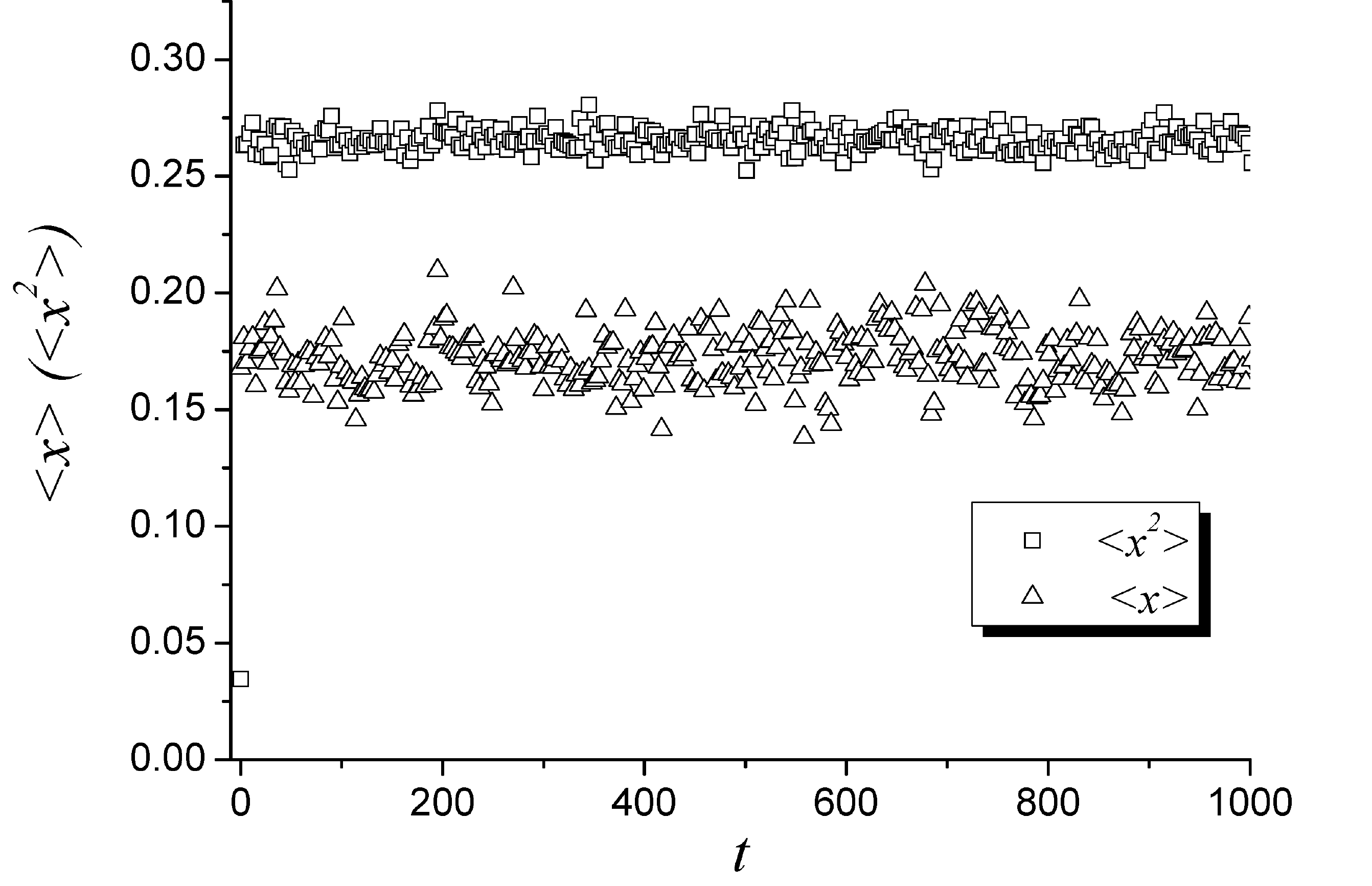} \ c)\includegraphics[
 width=40mm]{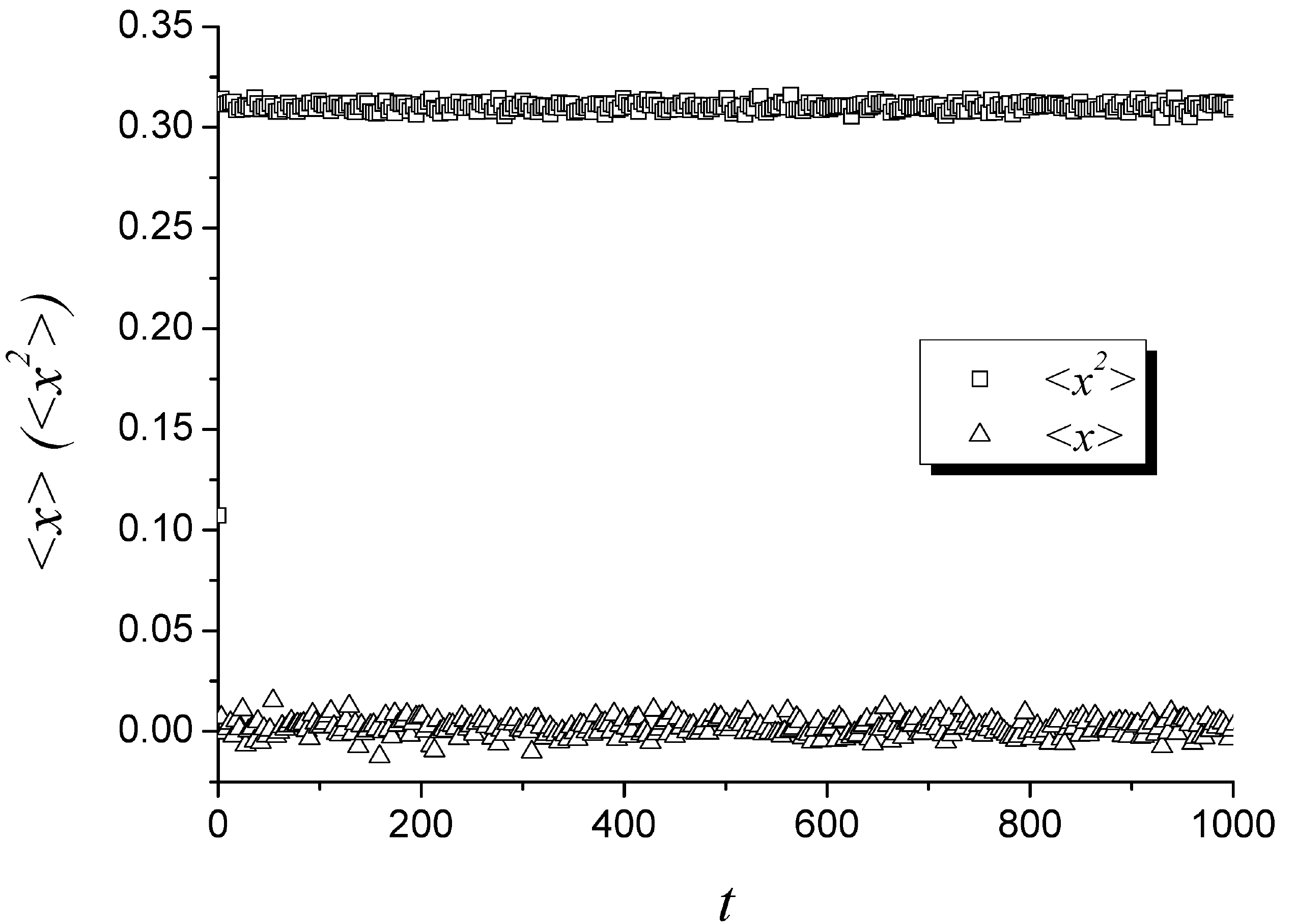}
 \caption{Evolution of first and second statistical moments at   $\sigma^2=0.2$ (a),
$\sigma^2=\sigma^2_0=1.2757$ (b) and $\sigma^2=2.0$ (c). Other parameters are:
$\varepsilon=0.2$, $\mu=0.5$, $\kappa=1.0$, $\beta=1.0$,
$\alpha=0.2$\label{mom_t}}
\end{figure}
\begin{figure}[!h]
\centering
 \includegraphics[
 width=80mm]{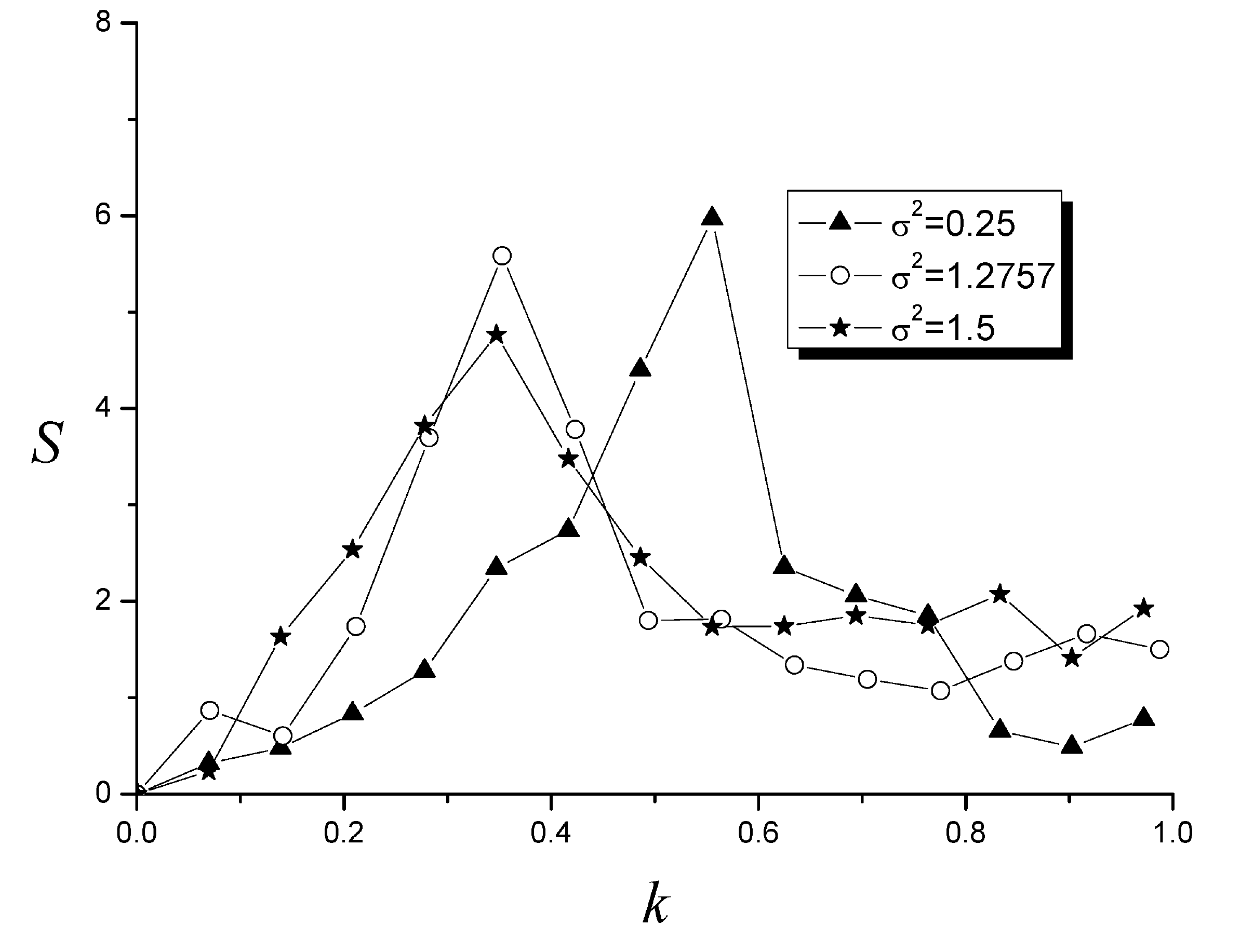}
 \caption{An averaged structure function at different values of the noise intensity at $t=4000$ (the model of lateral interactions): triangles  correspond to $\sigma^2=0.25$;
 circles correspond to $\sigma^2=\sigma^2_0$; stars relate to $\sigma^2=1.5$.  Other parameters are:
$\varepsilon=0.2$, $\mu=0.5$, $\kappa=1.0$, $\beta=1.0$,
$\alpha=0.2$\label{S(k)}}
\end{figure}
The stationary value of the first statistical moment $\langle
x(\mathbf{r},t)\rangle$ goes from initial one $\langle
x(\mathbf{r},0)\rangle\simeq x_-$ to a positive fixed value (see Fig.8a ). It
means that at small noise intensity the effective potential $U_{ef}(x)$ has a
broken symmetry: $U_{ef}(x_-)>U_{ef}(x_+)$. It leads to the effect when in a
course of evolution the value $\langle x\rangle$ goes to a positive constant.
At $\sigma^2=\sigma^2_0$ one has $U_{ef}(0)=U_{ef}(x_+)$. Then, starting from
the configuration $\langle x(\mathbf{r},0)\rangle= x_- >0$ with small
dispersion a picture kind of phase separation should be realized. Indeed,
despite the picture shown in Fig.7b is noisy with domains of two phases and
diffuse interfaces, the statistical average $\langle x\rangle$ does not change
its value, i.e. $\langle x(\mathbf{r},0)\rangle=\langle
x(\mathbf{r},t\to\infty)\rangle=x_-$ (see Fig.8b). Therefore, here we have a
situation when dynamics is conserved: $\int x(\mathbf{r},t){\rm
d}{\mathbf{r}}=const$, but there are no domain size growth law as it realizes
in phase separation processes. At $\sigma^2>\sigma^2_c$ (see Fig.7c) when the
effective potential has only minimum placed at $x=0$ the numerical solution
(see Fig.7c) shows massive structures with $\langle
x(\mathbf{r},t)\rangle\simeq 0.0$. Despite the system has trivial value for the
first statistical moment, the second moment is not trivial (see Fig.8). We
calculate the structure function at different values $\sigma^2$. The
corresponding dependencies are shown in Fig.9. It is seen that $S(k)$ has one
peak located at values $k$ determined by the linear stability analysis (see
Fig.3a) and depends on the noise intensity. With an increase in $\sigma^2$ the
peak smears, it means that patterns have more diffuse interface.

The bifurcation values for $\sigma^2$ illustrating a qualitative change of the
stationary distribution (see Fig.2a) can be calculated numerically. Due to we
can not investigate a form of the total stationary distribution functional
$\mathcal{P}_{st}[x]$ numerically, our analysis were performed for the quantity
$\delta=\langle x\rangle-x_{mp}$, where $x_{mp}$ is the most probable value
related to the position of the maximum for the distribution function $P_{st}$.
In such a case $\delta$ plays a role of a criterium for the quantitative change
of the stationary distribution form. Indeed, considering the interval
$\sigma^2\in[0, \sigma^2_s)$, one choose $x_{mp}=0$ that yields $\delta=\langle
x\rangle>0$. It means that the stationary distribution has a broken symmetry
with a large contribution related to values $x>0$. At $\sigma^2=\sigma^2_s$ the
stationary distribution has a double degenerated point located at $x=x_0=x_-$.
It leads to the fact that a dependence $\delta(\sigma^2)$ breaks at
$\sigma^2=\sigma^2_s$ and decreases with $\sigma^2$ growth; one has to take
into account the value $x_{mp}=x_- >0$ in the parameter $\delta$ calculation at
$\sigma^2\in [\sigma^2_s, \sigma^2_c)$. At $\sigma^2=\sigma^2_0$ one has
$\langle x\rangle=x_-$ and $\delta=0$. It means that the stationary
distribution is of a symmetrical form with respect to the central extremum
$x=x_-$. With further increase in the noise intensity the parameter $\delta$
becomes negative and changes its value abruptly at $\sigma^2=\sigma^2_c$ where
$x_{mp}=0$. The corresponding dependencies (analytical and numerical) are shown
in Fig.10. It is seen a good correspondence of analytical results with computer
simulations.

Previously, we have studied stochastic dynamics when the system evolves into
the stationary state, described by the functional $\mathcal{P}_{st}[x]\propto
\exp(-\mathcal{U}_{ef}[x]/\sigma^2)$. Let us consider the case when the system
relaxes according to the stationary distribution. To generate the field $x$
according to this distribution we can use the algorithm known variously as the
``Langevin method'' \cite{McKeown}. The principle idea lies in a supposition
that the distribution is realized, already. Then, the stochastic field $x$ can
be obtained as solution of an effective Langevin equation $
\partial_t x=-{\delta \mathcal{U}_{ef}[x]}/{\delta x}+\xi(\mathbf{r},t)$,
where $\xi(\mathbf{r},t)$ is a white additive noise with intensity $\sigma^2$.
Solutions of the deterministic equation $\partial_t x=-{\delta
\mathcal{U}_{ef}[x]}/{\delta x}$ at $t\to \infty$ allows to give the stationary
patterns at fixed noise intensities. Here one assumes that the system is
distributed according to an effective probability density functional
$\mathcal{P}_{ef}=\prod_i\left<\delta_i({\delta \mathcal{U}_{ef}[x]}/{\delta
x})\right>$, where average is taken over initial conditions, $i$ relates to
minima positions of the function $U_{ef}(x)$. The corresponding patterns at
different $\sigma^2$ are shown in Fig.11. It is seen that at small noise
intensity (Fig.11a) nuclei are formed, at intermediate values
($\sigma=\sigma^2_0$) a picture type of spinodal decomposition is realized (see
Fig.11b), at large $\sigma^2$ (see Fig.11c) linear defects of dislocations kind
are realized. Obtained pictures correspond to profiles of the concentration
$x_{st}$, shown in Fig.5: at small noise intensities the pointed profile is
observed, whereas at intermediate and large $\sigma^2$ a symmetrical form of
profiles are realized.
\begin{figure}
\centering
 \includegraphics[width=80mm
 ]{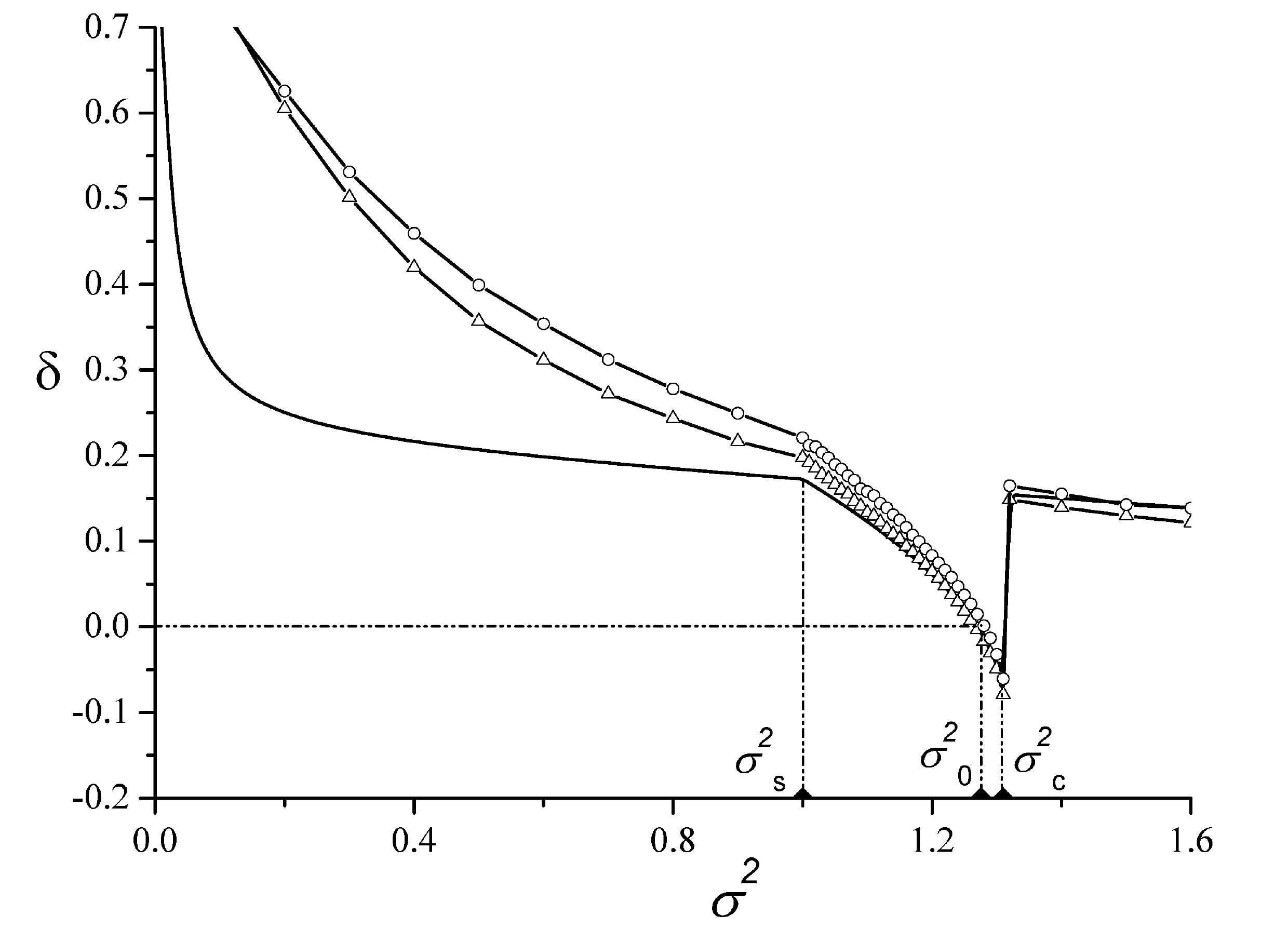}
\caption{The effective order parameter $\delta$ versus noise intensity: solid
line corresponds to analytical investigation; triangles correspond to
simulations on the lattice with linear size $N=96$; circles are related to
$N=192$. Other parameters are: $\alpha=0.2$, $\varepsilon=0.2$, $\mu=-0.5$,
$\beta=1.0$, $\kappa=1$. \label{delta}}
\end{figure}
\begin{figure}
\centering
 a)\includegraphics[width=40mm]{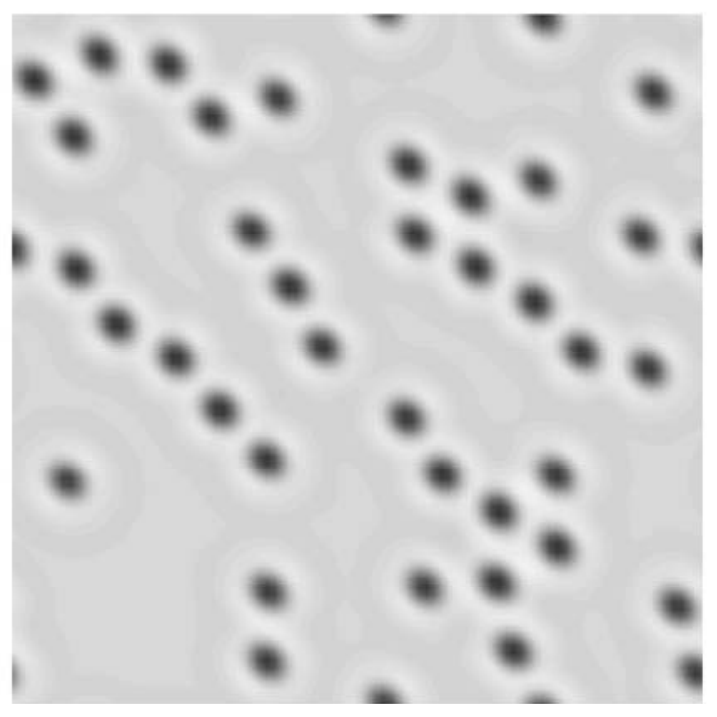}\ b)\includegraphics[width=40mm]{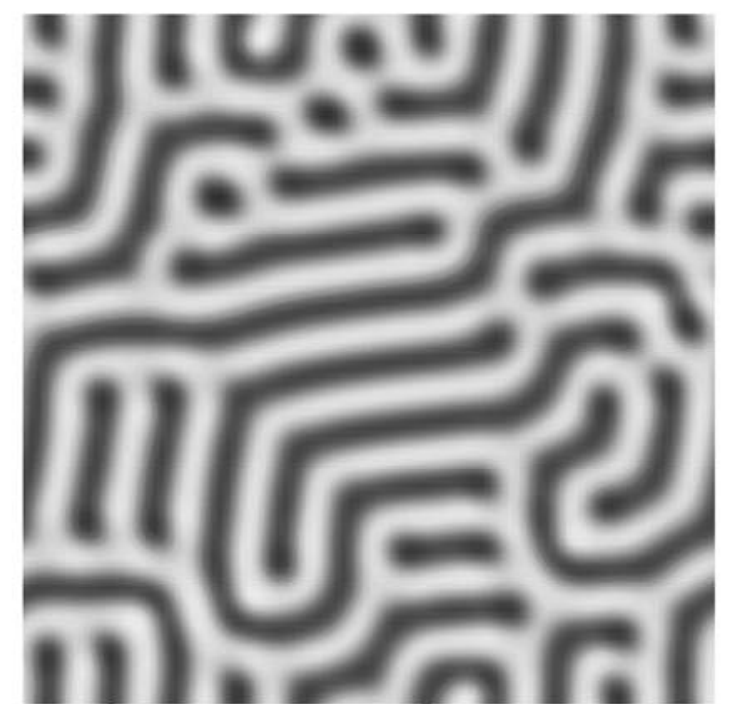}
 c)\includegraphics[width=40mm]{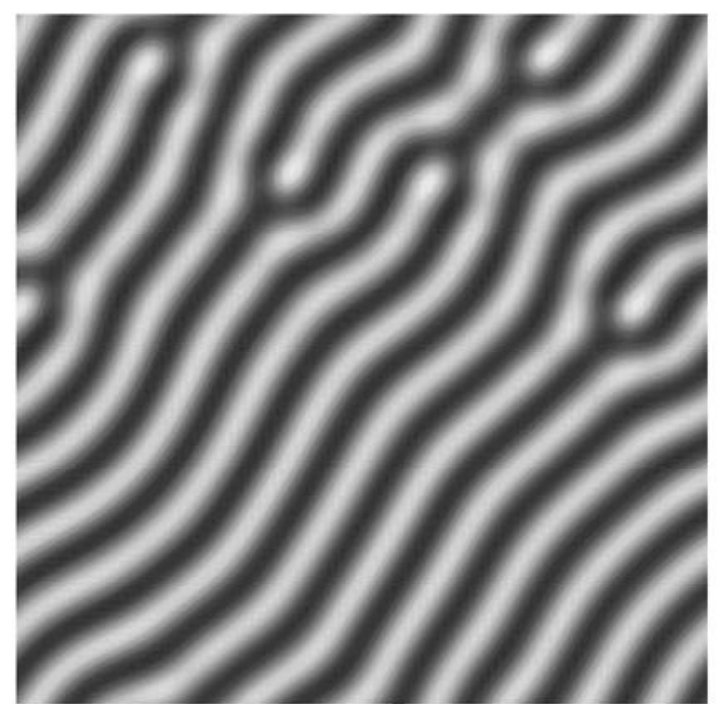}\\
\caption{Stationary patterns at $\sigma^2=0.2$ (a), $\sigma^2=\sigma^2_0$ (b)
and $\sigma^2=2$(c). Other parameters are: $\alpha=0.2$, $\varepsilon=0.2$,
$\mu=-0.5$, $\beta=1.0$, $\kappa=1$. \label{Stpt}}
\end{figure}
\section{Conclusions}
We have studied a simple deterministic model of reaction-diffusion systems
which can qualitatively describe stationary noise patterns. In order to study
noise induced mechanism for pattern formation we add a multiplicative noise in
a way to satisfy a fluctuation-dissipation relation. We have found that the
stationary distribution can be obtained exactly. Comparing noise induced
transitions picture and pattern formation processes it was shown that the
system follows the entropy driven mechanism by analogy with entropy driven
phase transitions theory.

We have explored early stages of the system evolution and found that at small
times the noise leads to instability of the null state. Considering the
stationary case we solve the variation problem when the stationary distribution
functional $\mathcal{P}_{st}[x]$ is maximized. Solutions of the problem show
that stationary patterns are formed in the vicinity of the maxima of the
corresponding stationary density function, ${P}_{st}(x)$. In linear stability
analysis in $\mathbf{r}$-space in the stationary case we have found that at
small noise intensities the system is characterized by unstable homogeneous
solutions related to the maxima positions of the function ${P}_{st}(x)$. With
an increase in the noise strength the spatial structures $x_{st}(\mathbf{r})$
are formed in the vicinity of the local maxima of ${P}_{st}(x)$. Analytically
and numerically we have obtained that at large noise intensity after a
bifurcation point (the stationary probability density function ${P}_{st}(x)$ is
of one maximum) the noise can sustain stationary patterns. The process of the
pattern formation is well described by the formalism of noise induced
transitions in zero-dimensional systems, whereas the obtained effect is a some
generalization of the noise induced transitions for extended systems. Our study
shows that at small noise intensity the system manifest a nucleation regime, at
fixed value of the noise strength a spinodal decomposition is realized, at
large noise the system exhibits strip patterns with liner defects. Strip
structures exist in the fixed interval of the noise intensity --- large
fluctuations destroy the patterns.

Our results are in a good correspondence with experimental data obtained for
patterns formation in polymer gels with chemical reactions (see
Ref.\cite{Katsuragi}). It was shown that in such the reaction-diffusion system
patterns can be formed only inside the fixed temperature interval
($10^0C<T<60^0C$). Author discusses diffusion-induced pattern formation
mechanism of volume phase transition with pattern formation. He shows that
straight strip patterns tend to appear in high temperature regime ($T\approx
57^0C$), and random patterns appear at low temperature ($T\approx 30^0C$).
Moreover, the size of patterns decrease in a power law form with an increase in
the temperature, $L\propto T^{-\beta/2}$, $\beta=1.01\pm0.03$. In our
calculations we have not used the corresponding chemical reactions, the local
dynamics were governed by the Schl$\ddot{{\rm o}}$gl model \cite{schlogl}, we
have used the model of non-Fickian transport with the field dependent diffusion
coefficient. The noise source satisfies the fluctuation-dissipation theorem and
has intensity related to the bath temperature. To compare our results with
above experimental investigations we need to stress that an actual noise
intensity values are $\sigma^2\in(\sigma^2_s,\sigma^2_{T0}]$. We have shown
that random stationary patterns are realized if
$\sigma^2_s<\sigma^2<\sigma^2_c$ (experimental photograph in
Ref.\cite{Katsuragi} is the same as we show in Fig.11b). At
$\sigma^2_c<\sigma^2_{T0}$ the strip patterns are realized (see Fig.11c), the
same picture was obtained in Ref.\cite{Katsuragi}. Moreover, we have calculated
the dependence $L\propto (\sigma^2)^{-\beta/2}$, and obtained
$\beta=3.2\pm0.02$. The qualitative correspondence in the wavelength dependence
is observed. The difference in the exponent $\beta$ values can be explained by
the fact that we have used another kind of the local dynamics model,
non-Fickian transport and white noise assumptions.

Obtained results can be applied to study patterns in adsorption/desorption
processes in metals \cite{ChemReact} deposition of a monolayer of molecules
\cite{Mik96} and in processes of microstructure transformations of materials
subject to intensive irradiation. One can await that formation of point defect
clusters and clusters of particles in the system kind of ``vacanies+particles''
where a phase decompositions and ``chemical transformations'' caused by
irradiation can be observed.

\end{document}